\documentclass[10pt,twocolumn]{IEEEtran}
\usepackage{amsfonts,amsmath,amsthm,latexsym,amssymb}
\usepackage{graphicx}
\usepackage{subfigure}
\usepackage{stfloats}
\usepackage{url}

\begin{document}

\title{An Efficient MAC Protocol with Selective Grouping and Cooperative Sensing in Cognitive Radio Networks}
\renewcommand{\baselinestretch}{1}

\author{Yi~Liu,
        Shengli~Xie,~\IEEEmembership{Senior Member,~IEEE,}
        Rong~Yu,
        Yan~Zhang,~\IEEEmembership{Senior Member,~IEEE,}
        and~Chau~Yuen~\IEEEmembership{Senior Member,~IEEE}

\thanks{This work is partially supported by National Natural Science Foundation of China (grant no.~U1035001,~U1201253,~ 61273192),
projects 217006/E20 funded by the Research Council of Norway, the
European Commission COST Action IC0902, IC0905 and IC1004, and the
European Commission FP7 Project EVANS (grant no. 2010-269323),
International Design Center (grant no.~IDG31100102 \& IDD11100101).}

\thanks{Yi~Liu (corresponding author) is with Guangdong University of Technology, Guangzhou,
China; Singapore University of Technology and Design, Singapore.
E-mail: liuii5115@yahoo.com.cn.}

\thanks{Shengli~Xie, and Rong~Yu are with Guangdong University of Technology, Guangzhou,
China. E-mail: shlxie@gdut.edu.cn, yurong@ieee.org.}

\thanks{Yan Zhang is with Simula Research Laboratory, Norway. E-mail:yanzhang@simula.no.}

\thanks{Chau~Yuen is with Singapore University of Technology and Design,
Singapore. E-mail: yuenchau@sutd.edu.sg.}}

\maketitle


\begin{abstract}


In cognitive radio networks, spectrum sensing is a crucial technique
to discover spectrum opportunities for the Secondary Users (SUs).
The quality of spectrum sensing is evaluated by both sensing
accuracy and sensing efficiency. Here, sensing accuracy is
represented by the false alarm probability and the detection
probability while sensing efficiency is represented by the sensing
overhead and network throughput. In this paper, we propose a
group-based cooperative Medium Access Control (MAC) protocol called
GC-MAC, which addresses the tradeoff between sensing accuracy and
efficiency. In GC-MAC, the cooperative SUs are grouped into several
teams. During a sensing period, each team senses a different channel
while SUs in the same team perform the joint detection on the
targeted channel. The sensing process will not stop unless an
available channel is discovered. To reduce the sensing overhead, an
SU-selecting algorithm is presented to selectively choose the
cooperative SUs based on the channel dynamics and usage patterns.
Then, an analytical model is built to study the sensing
accuracy-efficiency tradeoff under two types of channel conditions:
time-invariant channel and time-varying channel. An optimization
problem that maximizes achievable throughput is formulated to
optimize the important design parameters. Both saturation and
non-saturation situations are investigated with respect to
throughput and sensing overhead. Simulation results indicate that
the proposed protocol is able to significantly decrease sensing
overhead and increase network throughput with guaranteed sensing
accuracy.

\end{abstract}


\section{Introduction}

\IEEEPARstart{R}ecently, the explosive increase of wireless devices
and applications poses a serious problem of compelling need of
numerous radio spectrum. The problem is greatly caused by the
current fixed frequency allocation policy, which allocates a fixed
frequency band to a specific wireless system. On the contrary, a
recent report published by the Federal Communication Commission
(FCC) reveals that most of the licensed spectrum is rarely utilized
continuously across time and space \cite{FCC2}. In order to address
the spectrum scarcity and the spectrum under-utilization, Cognitive
Radio (CR) has been proposed to effectively utilize the spectrum
\cite{S. Haykin}-\cite{newadd2}. In the CR networks, the Secondary
(unlicensed) Users (SUs) are allowed to opportunistically operate in
the frequency bands originally allocated to the Primary (licensed)
Users (PUs) when the bands are not occupied by PUs. SUs are capable
to sense unused bands and adjust transmission parameters
accordingly, which makes CR an excellent candidate technology for
improving spectrum utilization.

Spectrum sensing is a fundamental technology for SUs to efficiently
and accurately detect PUs in order to avoid the interference to
primary networks. However, in CR networks, many unreliable
conditions \cite{Simon}-\cite{Jianghai1}, such as channel
uncertainty, noise uncertainty and no knowledge of primary signals,
will degrade the performance of spectrum sensing. Cooperative
sensing \cite{Jianghai2}-\cite{C.yuen}, has been studied extensively
as a promising alternative to improve sensing performance at both of
the physical (PHY) level and the medium access control (MAC) level.
The main interest of this paper is the cooperative sensing mechanism
at the MAC level, which performs sensing operations in two aspects:
1) assign multiple SUs to sense a single channel for improving the
sensing accuracy; 2) assign cooperative SUs to search for available
spectrums in parallel to enhance sensing efficiency.

The improvement of sensing accuracy is extensively treated in
\cite{Ganesan}-\cite{Zhi Quan}. The study in \cite{Ganesan} reports
a cooperative sensing approach through multi-user cooperation and
evaluates the sensing accuracy. The authors of \cite{J. Shen}
consider cooperative sensing by using a counting rule and derive
optimal strategies under both the Neyman-Pearson criterion and the
Bayesian criterion. The study in \cite{Zhi Quan} presents a new
cooperative wide-band spectrum sensing scheme that exploits the
spatial diversity among multiple SUs which also contributes to
improve the sensing accuracy. These studies have mainly focused on
improving sensing accuracy while sensing efficiency has been
ignored. The enhancement for sensing efficiency has been
investigated in \cite{Hang Su}\cite{Y.Liu}. The study in \cite{Hang
Su} introduces an opportunistic multi-channel MAC protocol which
integrate two novel cooperative sensing mechanisms, i.e., random
sensing policy and negotiation-based sensing policy. The latter
strategy assigns SUs to collaboratively sense different channels to
improve the sensing efficiency. For the sake of reducing sensing
overhead, the authors of \cite{Y.Liu} propose a multi-channel
cooperative sensing scheme, where the cooperative SUs are optimally
selected to sense the distinct channels at the same time for sensing
efficiency. These works assume that the sensing accuracy of one
channel by a single SU is completely true which is may not be
practical in real communication systems.

In addition, literatures above did not consider the design of the
cooperative MAC protocol for distributed networks and perform
theoretical analysis of sensing overhead and throughput. Hence, we
are interested in achieving both sensing accuracy and sensing
efficiency by introducing a cooperation protocol in MAC layer for CR
networks. Several cognitive MAC protocols have been proposed in the
literature to address various issues in CR network \cite{Hang
Su}\cite{J. Jia}\cite{H.B.Salameh}-\cite{Q.Chen}. However, these
protocols do not leverage the benefit of cooperation at MAC layer
for enhancing the sensing efficiency without degrading the sensing
accuracy.

In this paper, we propose a group-based cooperative MAC protocol
called GC-MAC. In GC-MAC, the cooperative SUs are grouped into
several teams. During a sensing period, each team senses a different
channel. The sensing process will not stop unless an available
spectrum channel is discovered. The purpose of team division has
twofold: 1) sensing a channel by several SUs for the improvement of
sensing accuracy; 2) finding more spectrum opportunities by sensing
distinct channels by different teams. As a consequence, multiple
distinct channels can be simultaneously detected within one sensing
period which leads to the enhancement of sensing efficiency.

To reduce the sensing overhead, we propose an SU-selecting algorithm
for GC-MAC protocol. In the SU-selecting algorithm, we selectively
choose the optimal number of the cooperative SUs for each team based
on the channel occupation dynamics in order to substantially reduce
sensing overhead. We analyze the sensing overhead and throughput in
the saturation and no-saturation network cases, respectively. In the
saturation networks, each SU always has data to transmit. In the
non-saturation networks, an SU may have an empty queue. In every
network case, we consider two types of channel conditions:
time-invariant channel and time-varying channel. In each condition,
the sensing overhead and the throughput are incorporated into an
achievable throughput maximization problem, which is formulated to
find the key design parameters: the number of the cooperative teams
and the number of SUs in one team. Furthermore, we present extensive
examples to demonstrate the sensing efficiency comparing with the
existing schemes and to show the determination of the crucial
parameters. Simulation results demonstrate that our proposed scheme
is able to achieve substantially higher throughput and lower sensing
overhead, comparing to existing mechanisms.

%
%
%
%
%
%
%


The remainder of this paper is organized as follows. In Section II,
the system  models are introduced. Section III reports our proposed
group-based MAC protocols for cooperative CR network. Section IV
introduces an SU-selecting algorithm for appropriately selecting the
cooperative SUs so as to reduce the sensing overhead. Then, we study
the sensing overhead and achievable throughput in the saturation and
non-saturation networks in Section V and Section VI, respectively.
Section VII evaluates the performance of the proposed GC-MAC
protocol based on our developed analytical models. Finally, we draw
our conclusions in Section VIII.

\section{Systems Models}

\subsection{Channel Usage Model}

We assume that each licensed channel alternates between ON and OFF
state, of which the OFF time is not used by PUs and hence can be
exploited by the SUs. Assume that the durations of the ON and the
OFF period are independently exponentially distributed. For a given
licensed channel, the duration of ON period follows an exponentially
distributed with parameter $\mu_{ON}$ and the duration of OFF period
with an exponentially distributed parameter $\mu_{OFF}$. We define
the channel availability as the normalized period which is available
for SUs. Let $p$ denotes the channel availability. Then, we have $
p=\frac{\mu_{ON}}{\mu_{OFF}+\mu_{ON}}$. Similar to \cite{Hang Su},
in this paper, we mainly consider that the licensed channels used by
the same set of PUs, i.e., the licensed channel availability
information sensed by each SUs is consistent among all SUs.

We consider two scenarios depending on the channel dynamics. The
first is the time-invariant channel with unchanged channel date rate
$R$. The throughput of the SU by using time-invariant channel only
depends on the constant data rate and the valid transmission time
$T_r$. The second type of channel is the Time-Varying Channel. The
Finite-State Markov Channel (FSMC) model is employed to model the
dynamics of the time-varying channel \cite{Markov channel}. The
dynamics of the time-varying channel is partitioned based on the
channel data rate. It is reasonable to employ the channel data rate
instead of Signal-to-Noise Ratio (SNR) which has been used in
conventional FSMC model. Since the channel data rate is closely
relevant to the application layer requirements and hence its usage
facilitates the construction of resource demands from an application
perspective. The set of the channel state is denoted as
$\mathbf{M}\equiv \{1,2,\dots, M\}$ with $\mid\!\mathbf{M}\!\mid=M
$. Let $c_i$ represents the channel state $i$ $(i \in \mathbf{M})$.
The state space is denoted as $\mathbf{S}\equiv \{c_i , i \in
\mathbf{M}\}$. Let $\pi_i$ $(i \in \mathbf{M})$ represents the
steady-state probability at state $c_i$. Then, the steady-state
probability can be solved using the similar technique in
\cite{Markov channel}. During data transmission within a frame, the
time-variation is slow enough that the channel data rate does not
change substantially. This assumption is acceptable due to the short
data transmission period within a frame and has been frequently
used, e.g. \cite{YingChang Liang} \cite{Hyoil Kim}.

\subsection{Energy Detection Model}

In order to discuss our problem, we employ Energy Detection \cite{A.
Ghasemi} as the spectrum sensing scheme. Both of the real-valued
signal model and the complex-valued signal model are used to
describe the received signal at the SU's receiver.

\subsubsection{Real-Valued Signal Model}
Let $t_s$ be the sensing time and $f_s$ be the sample frequency
during sensing time. We denotes $N$ as the number of samples in a
sensing period, i.e. $N=t_sf_s$. The received signal $r_k(n)$ at the
$n$th sample and the $k$th SU is given by,
\begin{equation}
\label{eqn1} r_k(n)=\left\{
\begin{array}{l}
w_k(n),\quad \quad \quad \quad H_0\\\nonumber
 s_k(n)+w_k(n),\quad H_1
\end{array}
\right.
\end{equation}
where $H_0$ represents the hypothesis that PUs are absent, and $H_1$
represents the hypothesis that PUs are present. $s_k(n)$ represents
the PU's transmitted signal which is assumed as a real-valued
Gaussian signal with mean zero and variance $\sigma_s^2$. $w_k(n)$
denotes a Gaussian process with mean zero and variance $\sigma_w^2$.

Let $e_k(r)$ denotes the test statistic of the $k$th SU. Then, we
have $ e_k(r)=\sum_{n=1}^N\mid r_k(n)\mid^2$. The detection and
false alarm probability of $k$th SU are given by,
\begin{equation}
\label{eqn3} P_d^k=Pr\left[e_k(r)>\lambda\mid H_1\right] \nonumber,
\quad P_f^k=Pr\left[e_k(r)>\lambda\mid H_0\right] \nonumber
\end{equation}
where $\lambda$ is a decision threshold of energy detector for a SU.

The test statistic $e(r)$ is known as Chi-square distribution with
$\frac{e(r)}{\sigma_w^2}\sim\chi_N^2$ under hypothesis $H_0$, and
$\frac{e(r)}{\sigma_s^2+\sigma_w^2}\sim\chi_N^2$ under hypothesis
$H_1$. However, if the number of samples is large, we can use the
Central Limit Theorem (CLT) to approximate the Chi-square
distribution by Gaussian distribution \cite{A. Ghasemi} under
hypothesis $H_z (z=0,1)$ with mean $\mu_z$ and variance $\sigma_z^2$
as,

\begin{equation}
\label{eqn5} \left\{
\begin{array}{l}
\mu_0=N\sigma_w^2, \quad \quad \quad \quad \quad \sigma_0^2=2N\sigma_w^4, \quad \quad \quad \quad \quad H_0\\
\mu_1=N(\sigma_s^2+\sigma_w^2), \quad \quad
\sigma_1^2=2N(\sigma_s^2+\sigma_w^2)^2,\quad  H_1. \nonumber
\end{array}
\right.
\end{equation}
Therefore, the probabilities $P_d^k$ and $P_f^k$ can be approximated
in terms of the $Q$ function is given by,

\begin{equation}
\label{eqn6}
P_d^k=Q\left(\frac{\lambda-N(\sigma_s^2+\sigma_w^2)}{\sqrt{2N}(\sigma_s^2+\sigma_w^2)}\right),
\nonumber \quad
P_f^k=Q\left(\frac{\lambda-N\sigma_w^2}{\sqrt{2N}\sigma_w^2}\right)
\end{equation}
where $Q(x)=\frac{1}{\sqrt{2\pi}}\int_x^\infty
e^{(-\frac{t^2}{2})}dt$.

\subsubsection{Complex-Valued Signal Model}

Considering the complex-valued signal model, the received signal
$r_k(n)$ at the $n$th sample and the $k$th SU can be given by,
\begin{equation}
\label{eqn1} r_k(n)=\left\{
\begin{array}{l}
w_k(n),\quad \quad \quad \quad H_0\\\nonumber
 h_k s_k(n)+w_k(n),\quad H_1
\end{array}
\right.
\end{equation}
where the channel coefficients $h_k$ is zero-mean, unit-variance
complex Gaussian random variables. $s_k(n)$ represents the PU's
transmitted signal which is assumed as a Gaussian signal with mean
zero and variance $\sigma_s^2$. $w_k(n)$ denotes a Gaussian process
with mean zero and variance $\sigma_w^2$.

The test statistic of the $k$th SU $ e_k(r)=\sum_{n=1}^N\mid
r_k(n)\mid^2$. The detection and false alarm probability of $k$th SU
are given by,
\begin{equation}
\label{eqn3} P_d^k=Pr\left[e_k(r)>\lambda_c \mid H_1\right]
\nonumber, \quad P_f^k=Pr\left[e_k(r)>\lambda_c \mid H_0\right],
\nonumber
\end{equation}
where $\lambda_{c}$ is a decision threshold of energy detector for a
single SU considering the complex-valued signal model. For a large
$N$, the distribution of $ e_k(r)$ can be approximated as Gaussian
distribution \cite{A. Ghasemi} with mean $\mu_z$ and variance
$\sigma_z^2$ under hypothesis $H_z (z=0,1)$ as,

\begin{equation}
\label{eqn5} \left\{
\begin{array}{l}
\mu_0=N\sigma_w^2, \quad \quad \quad \quad \quad \quad \quad \sigma_0^2=2N\sigma_w^4,\quad \quad \quad \quad \quad  \quad H_0\\
\mu_1=N(\mid h_k \mid ^2 \sigma_s^2+\sigma_w^2), \quad
\sigma_1^2=2N(\mid h_k \mid ^2 \sigma_s^2+\sigma_w^2)^2,  H_1
\nonumber
\end{array}
\right.
\end{equation}

Finally, we can obtain the probabilities $P_d^k$ and $P_f^k$ in
terms of the $Q$ function as

\begin{equation}
\label{eqn6} P_d^k=Q\left(\frac{\lambda-N(\mid h_k \mid
^2\sigma_s^2+\sigma_w^2)}{\sqrt{2N}(\mid h_k \mid ^2
\sigma_s^2+\sigma_w^2)}\right), \nonumber
P_f^k=Q\left(\frac{\lambda-N\sigma_w^2}{\sqrt{2N}\sigma_w^2}\right),
\end{equation}
where $Q(x)=\frac{1}{\sqrt{2\pi}}\int_x^\infty
e^{(-\frac{t^2}{2})}dt$.

\subsection{Counting Rule}

In order to improve sensing performance, an efficient fusion rule is
needed to make final decision to the availability of the channel.
Depending on every SUs' individual decision from one team, there are
three popular fusion rules: And-rule, OR-rule and Majority-rule
\cite{YingChang Liang}. And-rule mainly focuses on maximizing the
discovery of spectrum opportunities which are deemed to be exist if
only one decision says there is no PU. In OR-rule, as far as limit
the interference to the PU, the spectrum is assumed to be available
only when all the reporting decisions declare that no PU is present.
The last Majority-rule is based on majority of the individual
decisions. If more than half of the decisions declare the appearance
of primary user, then the final decision claims that there is a
primary user. Without loss of generality, we use the Majority-rule
in this paper with the assumption that all the individual decisions
are independent, and supposing that $P_d^{k}=P_d$ and $P_f^{k}=P_f$
\cite{YingChang Liang}. Then the joint detection probability and
false alarm probability by $j$ number of SUs are given by

\begin{equation}
\label{eqnpd} P_d(j)=\sum_{y=0}^{j-\lceil \frac{j}{2}\rceil}{j
\choose \lceil \frac{j}{2}\rceil+y}(1-P_d)^{j-\lceil
\frac{j}{2}\rceil-y}P_d^{\lceil \frac{j}{2}\rceil+y},
\end{equation}

\begin{equation}
\label{eqnpf}
 P_f(j)=\sum_{y=0}^{j-\lceil \frac{j}{2}\rceil}{j
\choose \lceil \frac{j}{2}\rceil+y}(1-P_f)^{j-\lceil
\frac{j}{2}\rceil-y}P_f^{\lceil \frac{j}{2}\rceil+y}.
\end{equation}

\section{GC-MAC: Group-based Cooperative MAC protocol}
In this section, we present the specifications of the proposed MAC
protocol, together with the group-based cooperative spectrum sensing
scheme and the SU-selecting algorithm. To describe our protocol
conveniently, we have the following assumptions:

\begin{itemize}
    \item Each SU is equipped with a single antenna which can not
    operate the sensing and transmission at the same time. According
    to this constraint, the sensing overhead caused by sensing is
    unavoidable and cannot be neglected in protocol design.
    \item A common control channel is available for all SUs to communicate at any time.
    \item An SUs can be assigned to perform
    cooperative sensing even when they have the packets to transmit.
\end{itemize}

A time frame of the secondary network operation is divided to three
phases: reservation, sensing and transmission. All SUs are
categorized into three types:

\begin{itemize}
    \item Source SU ($SU_s$): an SU that has data to transmit.
    \item
    Cooperative SUs ($SU_c$): SUs that are selected for
cooperative sensing.
    \item
    Destination SU ($SU_d$): an SU that receives the data packet from the source
SU.
\end{itemize}

\subsection{Reservation}

In GC-MAC,  any $SU_s$ entering the network first try to perform a
handshake with $SU_d$ on the control channel to reserve a data
channel. This allows the $SU_s$ and $SU_d$ to switch to the chosen
channel for data transmission. Here, we use R-RTS/R-CTS packets for
$SU_s$ and $SU_d$ to compete the data channel with other SUs. The
$SU_s$ will listen to control channel for a time interval $T$. If no
R-RTS/R-CTS is received or time $T$ is expired, the $SU_s$
participates in the reservation process. Otherwise, it will defer
and wait for the notification from the transmission pair or a
timeout. Whenever there is at least one packet buffered in the
queue, $SU_s$ sends reservation requirement to $SU_d$. Upon
receiving the requirement, $SU_d$ will reply and other SUs
overhearing these message exchanging cease their own sensing, and
wait for the notification from this transmission pair or a timer
expiration. When the sensing or cooperative sensing is finished,
other neighboring SUs start a new round of competition for the
control channel with a random backoff.

\subsection{Sensing}
\begin{figure}[htb]
\centerline{\includegraphics[width=7.5cm]{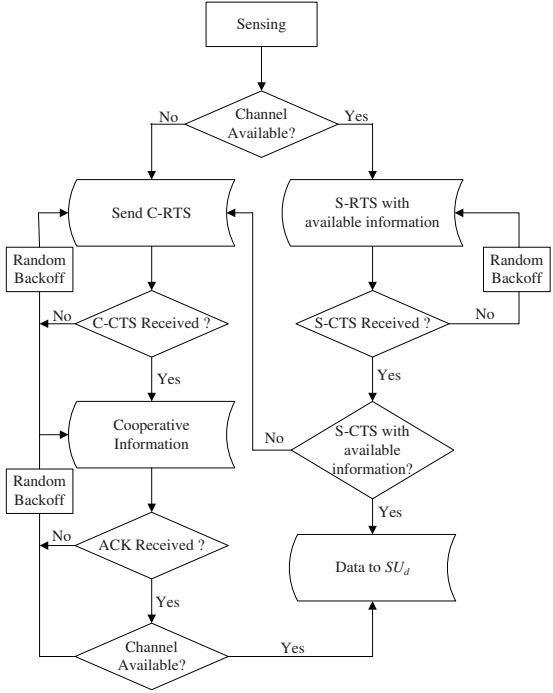}}
\caption{The sensing work flow of $SU_s$.} \label{SUs}
\end{figure}

\begin{figure}[htb]
\centerline{\includegraphics[width=9cm]{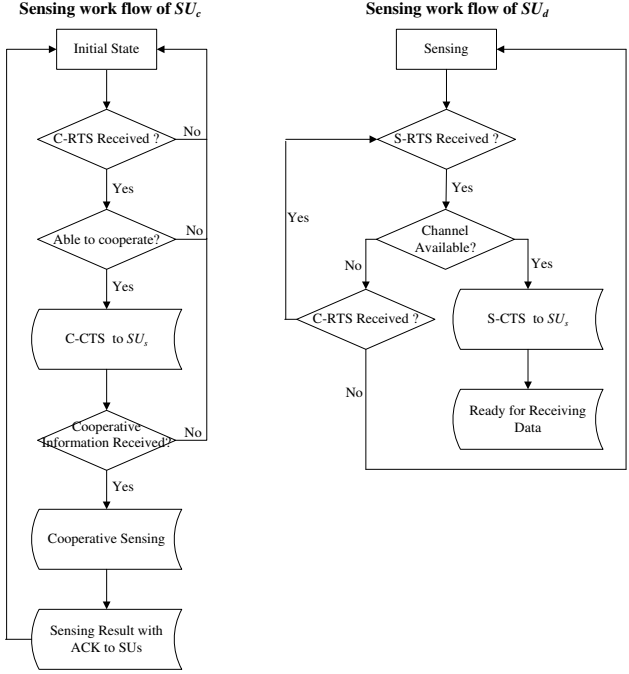}} \caption{The
sensing work flow of $SU_c$ and $SU_d$.} \label{SU}
\end{figure}

After reserving the data channel, $SU_s$ and $SU_d$ start to sense
the spectrum channel. In this phase, we use S-RTS/S-CTS packets for
spectrum sensing and negotiation between $SU_s$ and $SU_d$. In order
to indicate the mechanism of our scheme, the C-RTS/C-CTS packets are
included in the RTS/CTS model for $SU_c$ to acknowledge its
participation. Fig. \ref{SUs} shows the flowchart of the sensing
procedure of the source node $SU_s$. Fig. \ref{SU} shows the
flowchart of the sensing procedures of $SU_c$ and $SU_d$. In
particular, we provide the detailed description as follows.

 \textbf{Source SU} ($SU_s$)

1) $SU_s$ senses the channel to judge the availability of the
channel. If the channel is not occupied by a PU, $SU_s$ sends an
S-RTS packet to $SU_d$, including the availability information of
the detected channel. Otherwise, $SU_s$ sends the channel
unavailability information to $SU_d$.

2)  If an S-CTS packet from $SU_d$ is not heard after a CTS timer,
$SU_s$ should perform a random backoff, as if it encounters a
collision. If $SU_s$ receives the information of channel
availability from $SU_d$. $SU_s$ and $SU_d$ will start the
transmission phase (please refer to Section III.C). If $SU_s$
receives the information of channel unavailability from $SU_d$,
$SU_s$ will send C-RTS to the neighborhood of $SU_c$ and $SU_d$.

3) If $SU_s$ does not receive any feedback from $SU_c$, it then
sends cooperation requirement again after a random backoff. If the
feedback is successfully received, $SU_s$ counts the number of
$SU_c$ according to the SU-selecting algorithm (please refer to
Section IV). When the number of $SU_c$s satisfies the requirement of
the cooperative sensing, $SU_s$ stops sending cooperation
requirement to the neighborhood of $SU_c$ and divides the chosen
$SU_c$s into a number of teams.

4) $SU_s$ sends the cooperative information to the $SU_c$s and then
join the cooperative sensing with $SU_c$s. Such information includes
grouping information and the specific channels.

5) Upon receiving the sensing results, $SU_s$ should declare the
success of spectrum sensing and return to 1). Otherwise, $SU_s$
should perform a random backoff, and return to 4).

 \textbf{Cooperative SU} ($SU_c$)

1) Upon receiving the cooperation requirement, $SU_c$ sends feedback
to the source node $SU_s$ and waits for the cooperative information.

2) If the information for the cooperation is not received after a
CTS timer, $SU_c$ assumes that the information is lost and then
reverts to the original state. Otherwise, $SU_c$ starts the channel
sensing based on the cooperative information.

3) After the time duration $t_s$, $SU_c$ determines the PU's
activity on the detected channel and sends cooperation
acknowledgement to $SU_s$ with the sensing result.

     \textbf{Destination SU} ($SU_d$)

1) $SU_d$ senses the same channel with $SU_s$ in a synchronous way.
After the sensing time $t_s$, $SU_d$ makes the final decision about
the state (ON/OFF) of the channel, and waits for the sensing
requirement from the source node $SU_s$.

2) If the destination node $SU_d$ receives the sensing requirement
with the sensing result from the source node $SU_s$, it delivers the
sensing result back to $SU_s$. If the sensing result indicates that
the channel is available, $SU_d$ is ready for receiving data.
Otherwise, $SU_d$ waits for the cooperation requirement.

3) If cooperation requirement is received, $SU_d$ will join the
cooperative sensing and report the sensing results to $SU_s$. Then,
$SU_d$ returns to 2). If neither a sensing nor an cooperation
requirement is heard after a timer, $SU_d$ will go back to the
initial state.

\subsection{Transmission}

After the source node $SU_s$ and the destination node $SU_d$
successfully find an available channel, they begin to use the
channel to transmit data packets. Here, we use the T-RTS/T-CTS pair
to indicate the transmission process. Before starting the
transmission, $SU_s$ will send T-RTS to $SU_d$ for declaring the
beginning of transmission. Upon receiving this requirement, $SU_d$
replies T-CTS. If this feedback is received, $SU_s$ sends the data
packets to $SU_d$ and sets acknowledgment timeout. When the
acknowledgment from $SU_d$ arrives, $SU_s$ should declare the
transmission success over the control channel. This success
information ends the deferring of the neighboring SUs and starts a
new round of reservation. If acknowledgment is not received after an
acknowledgment timeout, $SU_s$ should perform a random backoff and
retransmit the data packets.

%

\section{Reducing Sensing Overhead Via SU-Selecting Algorithm}
In this section, we would like to reduce the sensing overhead by
introducing an SU-selecting algorithm. In this algorithm, we employ
the alternative pattern and the channel data rate of the SUs' used
channel as the cooperative SU's selection conditions.

\subsection{Channel Pattern for SUs}
Each channel alternates between state ON and state OFF which is
depending on the PUs' usage pattern. The channel that an SU uses may
be busy after a period $\tau$ based on the previous idle status.
During the busy period, the SUs are not allowed to access the
channels which are occupied by a PU. In this case, if these SUs are
selected for cooperative sensing, the overhead of cooperation can be
substantially reduced since sensing overhead is mainly incurred by
ceasing transmissions during the cooperative sensing period. Let
$I_\epsilon\in\{0,1\}$ represents the binary channel state of
channel $\epsilon$. $I_\epsilon=1$ refers to state ON and
$I_\epsilon=0$ refers to state OFF. Let $P_{I_\epsilon1}(\tau)$
denotes the transition probability that the $\epsilon$th channel
will be busy after $\tau$ seconds with the initial state
$I_\epsilon$. We can express the transition probability
$P_{01}(\tau)$ from channel state OFF to ON as \cite{A.Ghasemi}
\begin{equation}
\label{eqn3} P_{01}(\tau)= p-pe^{-(\mu_{OFF}+\mu_{ON})\tau}
\end{equation}
where $p$ is the channel availability.

It is shown that the $P_{01}(\tau)$ only relate with the most recent
channel state $I_\epsilon = 0$ and $\tau$, the time between the most
recent sensing and the current sensing. Considering that $\tau$ is
different among the channels, then $P_{01}(\tau)$ is accordingly
different with distinct SUs. In order to reduce the sensing
overhead, our goal is to select the cooperative SUs with the high
$P_{01}(\tau)$. In the following section, we first present the
optimal SUs-selecting algorithm in the time-invariant channel case.
Then, we derive the optimal selecting algorithm for the case where
the channel has time-varying feature.

\subsection{SU-Selecting Algorithm}

\subsubsection{Time-Invariant Channel Case}

A channel may stay at the idle state after $\tau$ seconds. The
sensing overhead is expected to be high if the SUs who used these
channels are chosen for cooperative sensing. Thereafter, in order to
reduce sensing overhead, we select the cooperative SUs in the
descending order of the probability $P_{01}(\tau)$. We can present
the SU-selecting algorithm as follows.
\begin{enumerate}
    \item
$SU_s$ delivers the Cooperative Sensing Request message (MSG-CSR) to
the neighboring $SU_c$s when a PU's activity is detected on a
channel.
    \item
The $k$th $SU_c$ calculates $P_{01}(\tau_k)$ where $\tau_k$
represents the time duration from the moment of the most recent
sensing to the moment of receiving MSG-CSR.
    \item
$SU_s$ selects the cooperative $SU_c$s according to the descending
order of $P_{01}(\tau_k)$.
\end{enumerate}
The probability $P_{00}(\tau_k)$ can be alternatively employed since
$P_{00}(\tau_k)=1-P_{01}(\tau_k)$. Hence, the SU-selecting algorithm
can obtain the same strategy if we choose the cooperative SUs in the
ascending order of the probability $P_{00}(\tau_k)$.

\subsubsection{Time-Varying Channel Case}

To reduce the sensing overhead, the SUs which have the highest
$P_{01}(\tau)$ should be selected for cooperation in the
time-invariant channel case. Here, the probability $P_{01}(\tau)$
represents the transition probability from state OFF to state ON.
However, this strategy may not be efficient in the time-varying case
where the channel data rate changes over the time. We choose the SUs
not only based on the probability $P_{01}({\tau})$ but also based on
the channel data rate of their used channels. The SUs' used channels
which have both the lowest channel data rate and the highest
$P_{01}(\tau)$ (or lowest $P_{00}(\tau)$) are selected to perform
sensing and search the available channels. As a consequence, in the
time-varying channel case, the SU-selecting algorithm can be
provided as follows.
\begin{enumerate}
    \item $SU_s$ delivers the Cooperative Sensing Request message (MSG-CSR) to the $SU_c$s when PU's
activity is detected on a channel.
    \item The $k$th $SU_c$ calculates $P_{00}(\tau_k)$, where $\tau_k$
represents the time duration from the moment of the most recent
sensing to the moment of receiving the message MSG-CSR.
    \item $SU_s$ multiplies $P_{00}(\tau_k)$ by the channel data rate $R_k$ of the $k$th SU's
    channel.
    \item $SU_s$ selects the cooperative SUs according to the ascending
    order
of $P_{00}(\tau_k)R_k$.
\end{enumerate}

\section{Analysis and Optimization for The Saturation Networks}
In this section, we will analyze the sensing overhead and throughput
in a saturation networks. Our objective is to find two key design
parameters: the number of cooperative teams and the number of SUs in
one team. In a saturation network, we consider the CR network
consisting of $C$ licensed channels and $K$ number of SUs. The set
of licensed channels is denoted as $\mathbf{C}\equiv\{1,2,\cdots,
C\}$ with $|\mathbf{C}|=C$. The set of SUs is denoted as
$\mathbf{K}\equiv\{1,2,\cdots, K\}$ with $|\mathbf{K}|=K$. We allow
the cooperative sensing scheme to choose a certain number of SUs
which are further divided into $U$ teams. Each team has $q$ $(q\geq
1)$ number of SUs and is assigned to sense a distinct channel during
each sensing period $t_s$. The relationship among the variables $K$,
$U$ and $q$ satisfies $Uq\leq K$.

\subsection{Time-Invariant Channel Case}

\subsubsection{Sensing Overhead}

We define $T_s$ as the total time duration spent by the $k$th
cooperative SU after $n_s$ number of the cooperative sensing. With
the proposed group-based sensing strategy, up to $U$ number of
channels can be detected in one sensing period. Hence, all channels
can be sensed completely within $\lceil C/U \rceil$ number of
sensing and the variable $n_s$ varies between 1 and $\lceil C/U
\rceil$. If the channels can be found after $n_s$ number of
cooperative sensing, the cooperative SUs can not transmit any
packets during $T_s=n_s t_s$ sensing periods. This operation is
unfortunately unavoidable in the cooperative sensing. Let
$o_{k}^{TI}$ denotes the sensing overhead caused by the $k$th
cooperative SU in the time-invariant situation. Then, we have
\begin{equation}
\label{eqn4} o_{k}^{TI}=\int_0^{T_s} R_k P_{00}(\tau_k)d\tau_k
\end{equation}
where $R_k$ denotes the channel data rate of the channel used by the
$k$th cooperative SU. Since the channel data rate is a constant in
the time-invariant channel case, we obtain sensing overhead as
\begin{equation}
\label{eqn5} o_{k}^{TI}=\int_0^{T_s} R P_{00}(\tau_k)d\tau_k.
\end{equation}

\subsubsection{Throughput}

Let $P_s$ represents the probability that a channel is successful
found. This is equal to the probability that a channel is available
and no false alarm is generated by $q$ number of cooperative SUs.
Then, we have $P_s=p\left[1-P_f(q)\right]$, where $p$ is the channel
availability and $P_f(q)$ is given by (\ref{eqnpf}). Let $u$ denotes
the number of available channels that are found in a cooperative
sensing. The probability distribution function of the random
variable $u$ is given by ${U \choose u}(1-P_s)^{U-u}P_s^u$. Then, we
can obtain the probability, $P_{av,1}^{TI}$, that the available
channels can be found in one cooperative sensing as
\begin{equation}
\label{eqn7} P_{av,1}^{TI}=\sum_{u=1}^U{U \choose
u}(1-P_s)^{U-u}P_s^u.
\end{equation}

With the proposed group-based sensing strategy, up to $U$ number of
channels can be detected in one sensing period. Hence, all channels
can be sensed completely within $\lceil C/U \rceil$ number of
sensing periods. We can then obtain the probability
$P_{av,n_s}^{TI}$ that an available channel is found after $n_s$
cooperative sensing.
\begin{equation}
\label{eqn8}
P_{av,n_s}^{TI}=\left(1-P_{av,1}^{TI}\right)^{n_s-1}P_{av,1}^{TI}.
\end{equation}

Let $T_r$ denotes the average transmission time for an SU using
discovered available channel. We can derive the throughput of an SU
by using this channel as follows
\begin{align}\notag
    \mathcal{T}^{TI}&=\sum_{n_s=1}^{\lceil C/U \rceil} P_{av,n_s}^{TI} T_r R
    \\\notag&= \sum_{n_s=1}^{\lceil C/U \rceil}  T_r R
    \left[1-\sum_{u=1}^U{U \choose u}(1-P_s)^{U-u}P_s^u\right]^{n_s-1}
    \\\label{eqn9}&
    ~~~~~~~~~~~\times \left[\sum_{u=1}^U{U \choose
    u}(1-P_s)^{U-u}P_s^u\right]
\end{align}
where $T_r=\int_0^\infty\mu_{OFF}e^{-\mu_{OFF}t}tdt=1/\mu_{OFF}$.

To determine the optimal value of $U$ and $q$, we introduce a new
term \textit{the achievable throughput}, which is defined as the
difference between sensing overhead and throughput. It is clear that
the achievable throughput is able to demonstrate the purely achieved
throughput after removing the penalty with respect to sensing
overhead. For this perspective, the concept is able to capture the
inherent tradeoff in the cooperative sensing.

Suppose that the available channel is discovered at the $n_s$th
detection by $U$ number of teams. We can obtain the total sensing
overhead $\mathcal{O}^{TI}$,
\begin{equation}
\label{eqn10} \mathcal{O}^{TI}=\sum_{n_s=1}^{\lceil C/U \rceil}n_s
P_{av,n_s}^{TI} q U o_{k}^{TI}.
\end{equation}

Our objective is to find the optimal $U$ and $q$ for the group
sensing in order to maximize the achievable throughput. The
optimization problem is formulated as
\begin{equation}
\label{eqn11}
\begin{split}
\max_{q,U} & \quad
\mathcal{G}^{TI}=\mathcal{T}^{TI}-\mathcal{O}^{TI} \\
\mathrm{s.t.} & \quad qU\leq K, \\
& \quad P_f (q)\leq P_{f, th},
\quad P_d (q)\geq P_{d, th},\\
\end{split}
\end{equation}
where $P_{f,th} $ and $P_{d,th} $ represent the threshold of the
false alarm probability and detection probability, respectively.
Based on the derived expression of $\mathcal{T}^{TI}$ and
$\mathcal{O}^{TI}$, the optimal number of cooperative teams and SUs
in one team can be determined by solving (\ref{eqn11}). Considering
the prohibitively high complexity of the optimization problem, we
have resorted to numerical methods to find the optimal result to
maximize the achievable throughput.

\subsection{Time-Varying Channel Case}

In this section, we will perform an analytical analysis on sensing
overhead and throughput in the time-varying channel case. It is
noteworthy that the analysis in the time-varying channel case is not
a trivial extension of the analysis in the time-invariant channel
case. On the one hand, the analysis in the time-invariant channel
case is necessary to provide an easy understanding of the SUs
cooperation behavior; and also the inherent trade-off between
throughput and sensing overhead. On the other hand, the time-varying
channel case is much more complicated than the time-invariant case
by considering the complex channel dynamics. The development of
sensing overhead and throughput is dependent on the channel dynamics
, which leads to new equations for channel data rate, sensing
overhead, throughput and hence achievable throughput in the
time-varying case.

\subsubsection{Sensing Overhead}

Based on the SU-selecting algorithm, we can analyze the sensing
overhead caused by the group-based sensing under the time-varying
channel case. Let $\overline{R}= [R_1,R_2,\cdots, R_M]$ represents
the channel data rate vector of length $M$. Without loss of
generality, we suppose $R_1< R_2<\dots< R_M$. Let $\overline{X}=[
X_1,\dots, X_K]$ be a random sample from $\overline{R}$ of length
$K$. Hereby, the vector $\overline{X}$ represents the specific value
of a parallel sensing and hence has length $K$ instead of $M$. Let
$X_k (k \in \mathbf{K})$ denotes the $k$th order statistics of the
sample. Employing order statistics theory \cite{Queuing System}, we
can derive the probability $\mathrm{Pr}\{X_k=R_n\}$ ($k \in
\mathbf{K}; n \in \mathbf{M}$) which shows that the $k$th SU's
channel data rate is equal to $R_n$. We suppose that there are
$(h-1)$ number of samples in $\overline{X}$ with the probability
$\mathrm{Pr}\{X_i<R_n\}$ $(1 \leq i \leq h-1; 1\leq h\leq k)$;
$(l-h+1)$ number of samples in $\overline{X}$ with the probability
$\mathrm{Pr}\{X_i=R_n\}$ $(1 \leq i \leq l-h+1; k\leq l\leq K)$; and
$(n-l)$ number of samples in $\overline{X}$ with the probability
$\mathrm{Pr}\{X_i>R_n\}$ $(1 \leq i \leq n-l)$.

The random variables $X_i$ are statistically independent and
identically distributed with the generic form $X$, we have
\begin{equation}\notag
    \mathrm{Pr}\{X<R_n\}=\sum_{R_i<R_n}\mathrm{Pr}\{X=R_i\}=\sum_{i=1}^{n-1}\pi_i.
\end{equation}
Since the $(h-1)$ samples could be any random samples from
$\overline{X}$, we obtain the probability of this case
${K\choose{h-1}}(\sum_{i=1}^{n-1}\pi_i)^{h-1}$.

For the probability $\mathrm{Pr}\{X_i=R_n\}$, we have
\begin{equation}\notag
    \mathrm{Pr}\{X=R_n\}=\pi_n.
\end{equation}
Since the number of $(l-h+1)$ samples could be any random samples
from the rest of $(K-h+1)$ samples of $X$, we obtain the probability
of this case as ${{K-h+1}\choose{l-h-1}}(\pi_n)^{l-h+1}$.
\begin{eqnarray}\notag
 \mathrm{Pr}\{X_i>R_n\}&=&1-\mathrm{Pr}\{X\leq R_n\}\nonumber\\
 &=&1-\sum_{R_i<R_n}\mathrm{Pr}\{X=R_n\}\nonumber\\
 &=&1-\sum_{i=1}^{n}\pi_i.\notag
\end{eqnarray}

Similarly, we obtain the probability of this condition as
${{K-1}\choose{K-l}}(1-\sum_{i=1}^{n}\pi_i)^{K-l}$. By summarizing
all possibilities, the probability $\mathrm{Pr}\{X_k=R_n\}$ is given
by (\ref{eqn22}). Then, the channel data rate of the selected SU,
denoted as $R_k$ ($k \in \mathbf{K}$), is given by
\begin{equation}
\label{eqn13} R_k=\sum_{n=1}^M \sum_{i=1}^{n}R_i
\mathrm{Pr}\{X_k=R_n\}.
\end{equation}

Let $o_k^{TV}$ denotes the sensing overhead caused by the
cooperative $\textrm{SU}_k$ after $n_s$ number of cooperative
sensing under the time-varying channel condition. We can obtain
\begin{equation}
\label{eqn14} o_k^{TV}=\int_0^{T_s}R_kP_{00}(\tau_k)d\tau_k
\end{equation}
where $T_s=n_s t_s$ denotes the time spent by the $k$th cooperative
SU after $n_s$ number of sensing.

\begin{figure*}[!t]
\normalsize 
\begin{align}\notag\label{eqn22}
\mathrm{Pr}\{X_k=R_n\}&\!=\!\sum_{l=k}^K\sum_{h=1}^k\left[{K\choose{h-1}}\left(\sum_{i=1}^{n-1}\pi_i\right)^{h-1}
{{K-h+1}\choose{l-h-1}}\left(\pi_n\right)^{l-h+1}
{{K-l}\choose{K-l}}\left(1-\sum_{i=1}^n\pi_i\right)^{K-l}\right]\\
 &\!=\!\sum_{l=k}^K\sum_{h=1}^k\left[\!\frac{K!}{(h-1)!(l-h+1)!(K-1)!}\!\left(\sum_{i=1}^{n-1}\pi_i\right)^{h-1}\!
\left(\pi_n\right)^{l-h+1}
\!\left(\!1\!-\!\sum_{i=1}^n\!\pi_i\!\right)^{K-l}\!\right]\!.
\end{align}
\hrulefill 
\end{figure*}

\subsubsection{Throughput}

Let $v$ represents the number of spectrum channels that are found in
a cooperative sensing. The probability density function (PDF) of the
random variable $v$ is given by ${U \choose v}(1-P_s)^{U-v}P_s^{v}$
where $P_s$ is given by (\ref{eqn6}). Let $P_{av}^{TV}$ denotes the
probability that an available channel can be found in one
cooperative sensing in the time-varying channel case. Then, we have
\begin{equation}
\label{eqn15} P_{av}^{TV}=\sum_{v=1}^U{U \choose
 v}(1-P_s)^{U-v}P_s^{v}.
\end{equation}

We need to find the available channel with the highest channel data
rate by the $U$ teams. We will select the channel that has the
highest channel data rate in these $v$ channels for the SU to
access. Let $R_m$ $(1 \leq m \leq M)$  denotes the highest channel
rate in these $v$ number of channels. It is noteworthy that the
subscript $m$ in $R_m$ represents the index of channel data rate,
which ranges from 1 to $M$. Let $P_{rate,v}$ denotes the probability
that there are channels whose maximum rate is no lower than $R_m$
$(1 \leq m \leq M)$ in the founded $v$ channels. Then, we have
\begin{equation}
\label{eqn16} P_{rate,v}=
\left(\sum_{i=1}^m\pi_i\right)^{v}{\left[1-\left(\sum_{i=1}^{m-1}\pi_i\right)^{v}\right]}.
\end{equation}

Conditioning on all possibilities on the random variable $v$, we
obtain the probability $P_{rate}$ that there are channels whose
maximum rate is no lower than $R_m$ $(1 \leq m \leq M)$
\begin{equation}
\label{eqn17} P_{rate}=\sum_{v=1}^U{U \choose
 v}(1-P_s)^{U-v}P_s^{v}
P_{rate,v}.
\end{equation}

We obtain the probability $P_{maxrate}$ that $R_m$ is the maximal
channel data rate from all discovered available channels.
\begin{equation}
\label{eqn18}
P_{maxrate}=\left(1-P_{av}^{TV}\right)^{n_s-1}P_{rate}.
\end{equation}

With the proposed sensing strategy, each sensing period may find up
to $U$ number of channels. Hence, all channels can be sensed
completely within $\lceil C/U \rceil$ number of sensing periods. We
can derive throughput of the SU by using this channel as
\begin{align}\notag
    \mathcal{T}^{TV}&=\sum_{n_s=1}^{\lceil C/U \rceil} \sum_{m=1}^M P_{maxrate} T_r R_m
    \\\notag&=\sum_{n_s=1}^{\lceil C/U \rceil}
    \sum_{m=1}^M  T_r R_m \left(1-P_{av}^{TV}\right)^{n_s-1}
\\\label{eqn19}&\sum_{v=1}^U{U \choose
 v}(1-P_s)^{U-v}P_s^{v}
\left(\sum_{i=1}^m\pi_i\right)^{v}{\left[1-\left(\sum_{i=1}^{m-1}\pi_i\right)^{v}\right]}
\end{align}
where $T_r=\int_0^\infty\mu_{OFF}e^{-\mu_{OFF}t}tdt=1/\mu_{OFF}$.

We formulate the achievable throughput optimization problem by
considering both throughput and sensing overhead in the time-varying
channel condition. The total sensing overhead $\mathcal{O}^{TV}$ is
given by
\begin{equation}
\label{eqn20} \mathcal{O}^{TV}=\sum_{n_s=1}^{\lceil C/U \rceil} n_s
q U P_{av}^{TV} o_{k}^{TV}.
\end{equation}

Consequently, the achievable throughput maximization problem in the
time-varying channel case is formulated as
\begin{equation}
\label{eqn21}
\begin{split}
\max_{q,U} & \quad
\mathcal{G}^{TV}=\mathcal{T}^{TV}-\mathcal{O}^{TV} \\
\mathrm{s.t.} & \quad qU\leq K, \\
& \quad P_f (q)\leq P_{f, th},
\quad P_d (q)\geq P_{d, th},\\
\end{split}
\end{equation}
where $\mathcal{T}^{TV}$ and $\mathcal{O}^{TV}$ are given by
(\ref{eqn19}) and (\ref{eqn20}), respectively. By solving
(\ref{eqn21}), we can find the optimal $U$ and $q$ for the group
sensing in order to maximize the achievable throughput.

\section{Analysis and Optimization for The Non-Saturation Networks}
In this section, we will derive the sensing overhead and throughput
in the non-saturation networks. Suppose that an SU may have an empty
queue. In this network, we consider a discrete-time queue with an
infinite capacity buffer for the queuing behavior of an SU. The
packets arrival of the SUs is assumed to be a Poisson process with
arrival rate $\lambda_{pac}$. The packets are served on a First-In
First-Out (FIFO) basis. The service time of each packet is modeled
as identically distributed nonnegative random variables, denoted as
${\chi}_n (n\ge1)$, whose arrival process is independent to each
another. The similar assumption has been frequently used in the
literature, e.g \cite{Hang Su}, \cite{H.Chhaya}. Let $F(t)$ denotes
the service time Cumulative Distribution Function (CDF) with mean
$0<1/\mu=\int_0^{\infty}t dF(t)$. Let $\rho$ represents the traffic
load and it is given by $ \rho = \frac{\lambda_{pac}}{\mu}$. For a
practical system, the traffic load is less than 1, i.e. $\rho<1$.

Similar to saturation network, we still consider the CR network
consisting of $C$ licensed channels and $K$ number of SUs. The
cooperative SUs are divided into $U$ teams. Each team has $q$
$(q\geq 1)$ number of SUs. Each team is assigned to sense a distinct
channel during each sensing period $t_s$. The relationship among the
variables $K$, $U$ and $q$ also satisfies $Uq\leq K$. Next, we will
formulate the throughput maximization problem with time-invariant
and time-varying channel, respectively.

\subsection{Time-Invariant Channel Case}

Since the channel data rate will not change with the time in
time-invariant channel case. The packet service time is a constant,
which means we are able to employ the single-server queuing model,
$M/D/1$, to evaluate the group sensing scheme with time-invariant
channel.

Based on the result of \cite{Queuing System}, the variance of
service time $E(\chi^2)=0$ in the $M/D/1$ model. Let
$\overline{N}^{TI}_q$ denotes the average number of packets in a
queue for time-invariant channel case. Then, we have

\begin{equation}
\label{eqn35}
\overline{N}^{TI}_q=\sum_{v=1}^{\infty}vp_{v+1}=\frac{\rho^2}{2(1-\rho)}.
\end{equation}

\subsubsection{Sensing Overhead}
To reduce the sensing overhead, we still select $qU$ SUs that have
the lowest channel data rate and least $P_{00}(t)$ among $K$ SUs in
the non-saturation network. As explained, each group sensing can
sense $U$ number of channels. Hence, all channels can be sensed
completely within $\lceil C/U \rceil$ number of group sensing. Let
$N_{sense}^{TI, n_s}$ be the total number of packets that can be
transmitted in the $n_s$ number of group sensing by the $qU$ sensing
SUs if they are not participating the group-based cooperative
sensing. $N_{sense}^{TI, n_s}$ is given by
\begin{equation}
\label{eqn37} N_{sense}^{TI, n_s}=\min\left\{n_s
qU\overline{N}^{TI}_q,(qUT_sR_{use})/l\right\}; 1 \leq n_s \leq
\lceil C/U \rceil
\end{equation}
where $R_{use}$ denotes the channel data rate of the using channel,
$l$ denotes the length of a packet, $T_s=t_sn_s$ and
$\overline{N}^{TI}_q$ is given by (\ref{eqn35}).

Suppose that the available channel is discovered at the $n_s$th
detection by $U$ number of teams in non-saturation network. Then, in
a time-invariant channel case, we can obtain the total sensing
overhead $\mathcal{O}^{TI}_{nonsat}$,
\begin{equation}
\label{eqn38} \mathcal{O}^{TI}_{nonsat}=\sum_{{n_s}=1}^{\lceil
C/U\rceil}P_{av,n_s}^{TI}N_{sense}^{TI, n_s} l
\end{equation}
where $P_{av,n_s}^{TI}$ is given by (\ref{eqn8}).
\subsubsection{Throughput}

Let $T_r$ denotes the average transmission time for an SU using
discovered available channel. In the time-invariant channel case,
the average number of packets that SUs send during $T_r$ at the
equilibrium state is given by

\begin{equation}
\label{eqn34}
N_D^{TI}=\min\left\{\overline{N}_q^{TI},(T_rR_{use})/l\right\}
\end{equation}
where $T_r=\int_0^\infty\mu_{OFF}e^{-\mu_{OFF}t}tdt=1/\mu_{OFF}$.

With the proposed sensing strategy, each sensing period may find up
to $U$ number of channels. Hence, all channels can be sensed
completely within $\lceil C/U \rceil$ number of sensing periods.
Hence, we can derive the throughput of an SU by using the discovered
available channel as follows.
\begin{equation}
\label{eqn36} \mathcal{T}^{TI}_{nonsat}=\sum_{{n_s}=1}^{\lceil
C/U\rceil} P_{av,n_s}^{TI}N_D^{TI}l
\end{equation}
where the item $P_{av,n_s}^{TI}$ is given by (\ref{eqn15}).

In terms of the achievable throughput maximization, we formulate the
following problem

\begin{equation} \label{eqn39}
\begin{split}
\max_{q,U} & \quad
\mathcal{G}^{TI}_{nonsat}=\mathcal{T}^{TI}_{nonsat}-\mathcal{O}^{TI}_{nonsat} \\
\mathrm{s.t.} & \quad qU\leq K, \\
& \quad P_f (q)\leq P_{f, th}, \quad P_d (q)\geq P_{d, th}.
\end {split}
\end{equation}

\subsection{Time-Varying Channel Case}

Considering the time-varying channel case, the channel data rate may
vary from time slot to time slot. This alternative indicates that an
SU's capacity is a random variable. Following this reasoning, we can
use the M/G/1 queuing model.

\subsubsection{Sensing Overhead}
 Since the service time of each packets
depends on the channel data rate, we can express the CDF $F(t)$ as
\begin{equation}\label{eqn40}
F(t)=l/R_i(t)
\end{equation}
where $R_i(t)$ denotes the channel data rate of the $i$th channel
state at the $t$th time slot. Let $\overline{N}^{TV}_q$ denotes the
average number of packets in a queue for time-varying channel case.
Then, we have

\begin{equation}
\label{eqn41}
\overline{N}_q^{TV}=\sum_{v=1}^{\infty}vp_{v+1}=\frac{\lambda^2E(\chi^2)+\rho^2}{2(1-\rho)}
\end{equation}
where $E(\chi^2)=\int_0^{\infty}t^2 dF(t)$.

In the time-varying channel case, let $N_{sense}^{TV,n_s}$ be the
total number of packets that can not be transmitted by the $qU$
cooperative SUs in $n_s$ number of group sensing.
$N_{sense}^{TV,n_s}$ is given by
\begin{equation}
\label{eqn43} N_{sense}^{TV,n_s}=\min\left\{n_s qU{\overline
N}_q^{TV},(qUT_sR_{use})/l\right\}; 1 \leq n_s  \leq \lceil C/U
\rceil
\end{equation}
where $T_s=t_sn_s$ and $\overline {N}_q^{TV}$ is given by
(\ref{eqn41}).

Then, in a time-varying channel case, the total sensing overhead for
discovering an available channel can be obtained as follows
\begin{equation}
\label{eqn44} \mathcal{O}^{TV}_{nonsat}=\sum_{n_s=1}^{\lceil
C/U\rceil}P_{av}^{TV} N_{sense}^{TV,n_s} l.
\end{equation}

\subsubsection{Throughput}
We use $T_r$ to denote the average transmission time for an SU using
discovered available channel in the time-varying channel case. Then,
the average number of packets that SUs send during $T_r$ is given by

\begin{equation}
\label{eqn34}
N_D^{TV}=\min\left\{\overline{N}_q^{TV},(T_rR_{use})/l\right\}
\end{equation}
where $T_r=\int_0^\infty\mu_{OFF}e^{-\mu_{OFF}t}tdt=1/\mu_{OFF}$.

The proposed sensing strategy may find up to $U$ number of channels
during each sensing period. All channels can be sensed completely
within $\lceil C/U \rceil$ number of sensing periods. Suppose that
the available channel can be found after $n_s$ number of group
sensing, we can obtain the throughput of an SU by using discovered
available channel in the time-vary channel case.
\begin{equation}
\label{eqn42} \mathcal{T}^{TV}_{nonsat}=\sum_{n_s=1}^{\lceil
C/U\rceil} \sum_{m=1}^MP_{maxrate}N_D^{TV}l
\end{equation}
where the item $P_{maxrate}$ is given by (\ref{eqn18}).

Finally, we formulate the following problem in terms of achievable
throughput maximization

\begin{equation}
\label{eqn45}
\begin{split}
\max_{q,U} & \quad
\mathcal{G}^{TV}_{nonsat}=\mathcal{T}^{TV}_{nonsat}-\mathcal{O}^{TV}_{nonsat} \\
\mathrm{s.t.} & \quad qU\leq K, \\
& \quad P_f (q)\leq P_{f, th}, \quad P_d (q)\geq P_{d, th}.
\end {split}
\end{equation}

Considering the complexity of the optimization problems, we still
use numerical methods to find the optimal result to maximize the
achievable throughput in non-saturation network. The optimal results
are provided in the following section under time-invariant and
time-varying channel condition, respectively.

\section{Simulation Results}
In this section, we demonstrate the performance of the proposed
GC-MAC in CR networks. The network consists of total $C=10$ licensed
channels. The channel parameter of the OFF period $\mu_{OFF}=1/100$.
We concentrate on the low SNR situation, the SNR threshold for a PU
at the tagged SU is $\gamma=-10dB$. The channel bandwidth is 1 MHz
and the target probability of detection $P_d=0.9$ which is a
important parameter used by 802.22 standard \cite{802.22}. The
length of RTS/CTS packets and sensing period are 40Bytes and $1ms$,
respectively. Considering the time-varying channel case, the number
of channel data rate state is $M=10$. Accordingly, the channel data
rate of each channel ranges between $0.1 MB/s-1 MB/s$, which
decreases or increases its value by $10(\%)$ once every $5$ms.

\begin{table}[htb]
\caption{The achievable saturation throughput with different
combination of $U$ and $j$ in time-invariant channel case.}
\centerline{\includegraphics[width=8cm]{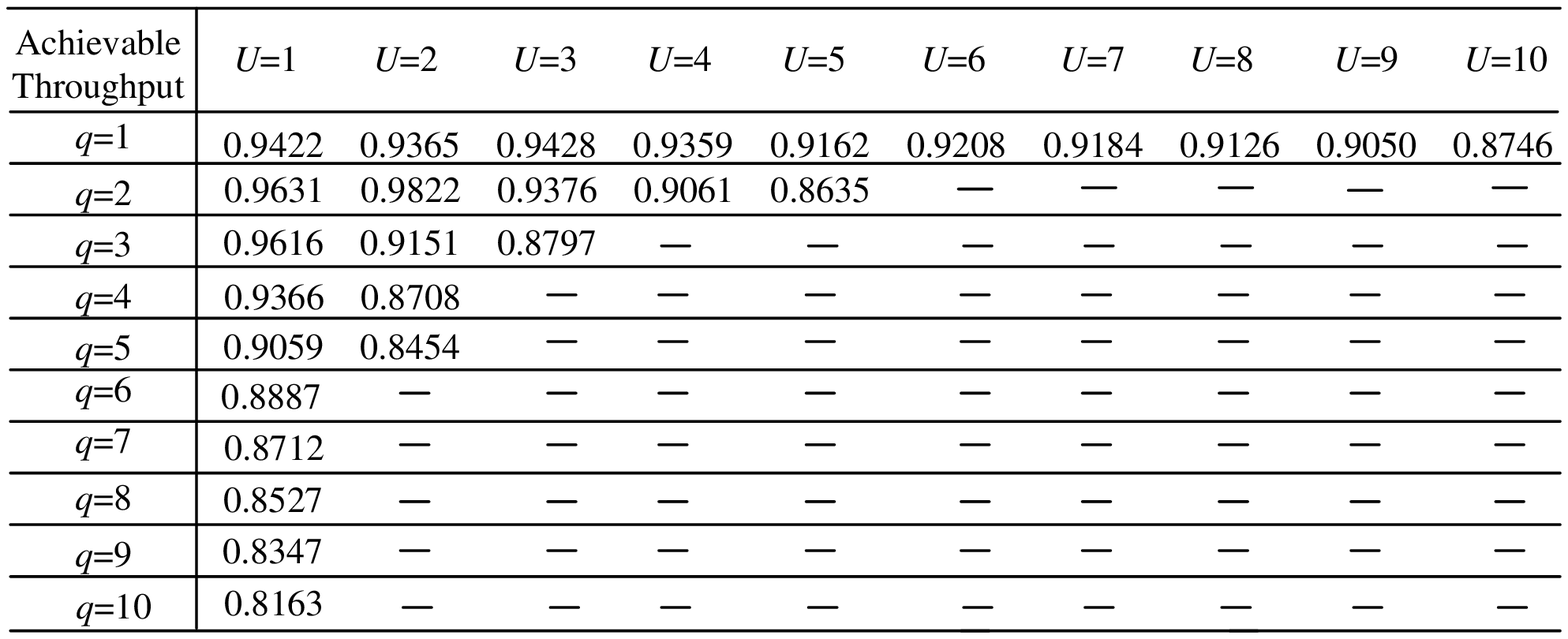}}
\label{table1}
\end{table}

\begin{table}[htb]
\caption{The achievable saturation throughput with different
combination of $U$ and $j$ in time-varying channel case.}
\centerline{\includegraphics[width=8cm]{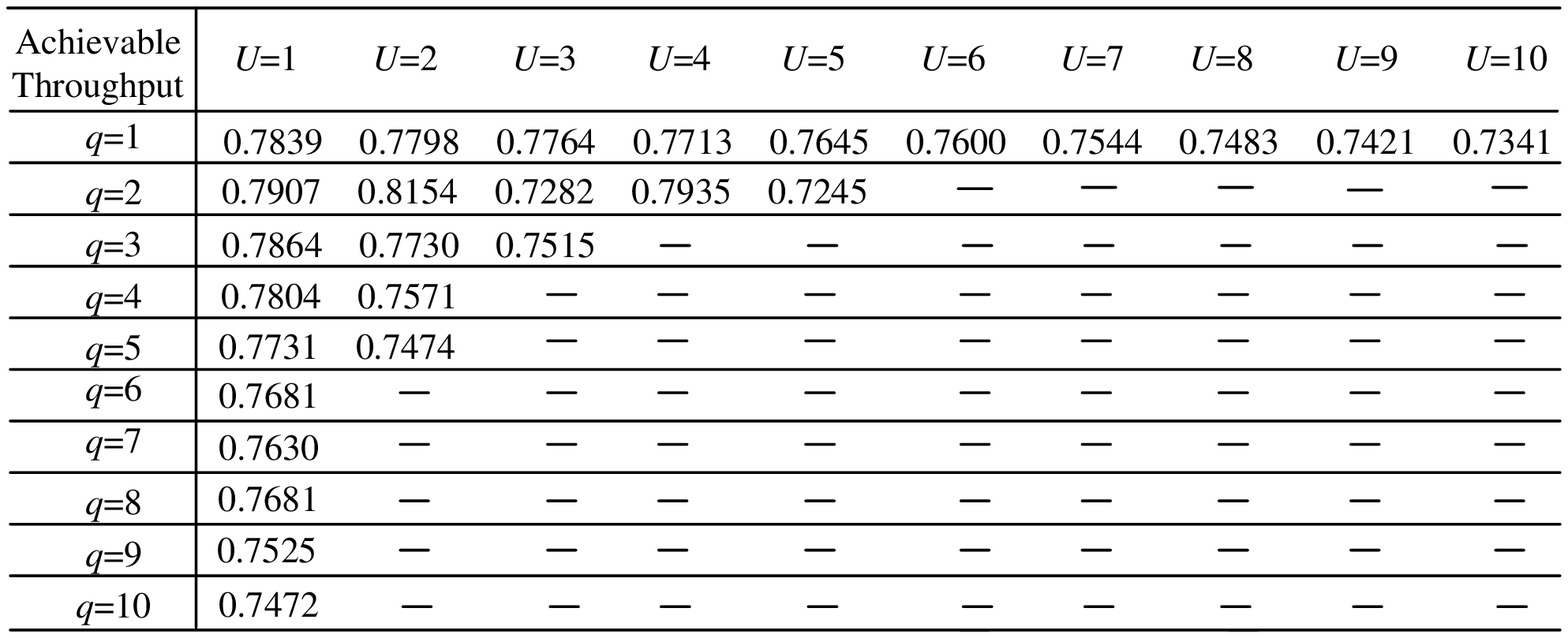}}
\label{table2}
\end{table}

Table.\ref{table1} shows the impacts of the number of cooperative
teams and the number of SUs in one team on the achievable saturation
throughput in the time-invariant channel situation. In these
examples, the channel availability $p$ is set as 1/2. We can
determine the optimal achievable throughput by choosing appropriate
parameters. From Table.\ref{table1}, we observe that the achievable
throughput is maximized as 0.9822. In the time-varying channel case,
Table.\ref{table2} shows the achievable saturation throughput that
the maximal value is 0.8154. The saturation throughput in the
time-varying case is lower than that in the time-invariant case.
This is expected since the channel data rate may be reduced in the
time-varying condition due to fading and signal variation.
Similarly, we can obtain the maximal non-saturation throughput in
the time-invariant channel case and the time-invariant channel case
as 0.9107 and 0.8095, respectively.

\subsection{Achievable Throughput}


We compare our GC-MAC which uses group-based cooperative sensing
scheme (GCSS) with accuracy priority cooperative sensing scheme
(ACSS) \cite{J. Shen} and efficiency priority cooperative sensing
scheme (ECSS) \cite{Hang Su}. In the scheme ACSS, every cooperative
SU monitors a single channel during each sensing period. The main
focus of this scheme is to improve sensing accuracy of a PU's
activity. In the scheme SCSS, the cooperative SUs are assigned to
sense different channels simultaneously for the sensing efficiency
enhancement. This sensing operation assumes that the sensing of each
channel by a single SU is accurate, which however may be difficult
to achieve in practical CR networks.


\subsubsection{Time-Invariant Channel Case}

\begin{figure}[htb]
\centerline{\includegraphics[width=7.5cm]{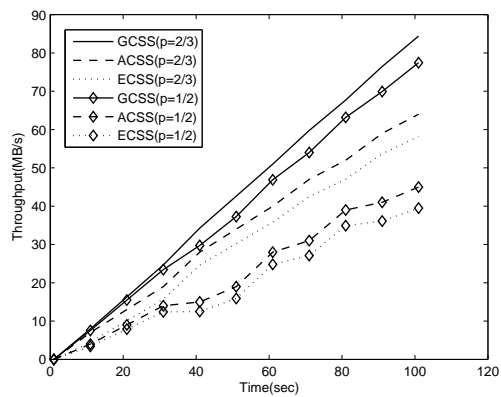}}
\caption{The saturation throughput in the time-invariant channel
case with different $p$.} \label{TH_INVSAT}
\end{figure}

Fig. \ref{TH_INVSAT} shows the throughput comparison among GCSS,
ACSS and ECSS in the time-invariant channel case when $p=2/3$ and
$1/2$. In this example, the sensing accuracy requirement is set as
$P_{f,th}=0.05$. It is observed that the achievable throughput in
all three schemes increases with higher channel availability $p$,
which is intuitively understandable. The result indicates that GCSS
is able to achieve much higher throughput than ACSS and ECSS. This
is because GCSS is able to search and find more spectrum
opportunities. When the number of the cooperative SUs becomes
larger, there is higher chance to find the available channels which
leads to less sensing overhead. In addition, ECSS uses all SUs to
sense different channels, which causes a less sensing accuracy of
single channel and leads to lower throughput. Comparatively, the
proposed GCSS chooses the optimal number of teams and the number of
SUs in each team. In this case, sensing overhead is significantly
reduced and throughput increases. As a consequence, our proposed
GCSS is able to achieve high sensing efficiency with low sensing
overhead.

\begin{figure}[htb]
\centerline{\includegraphics[width=7.5cm]{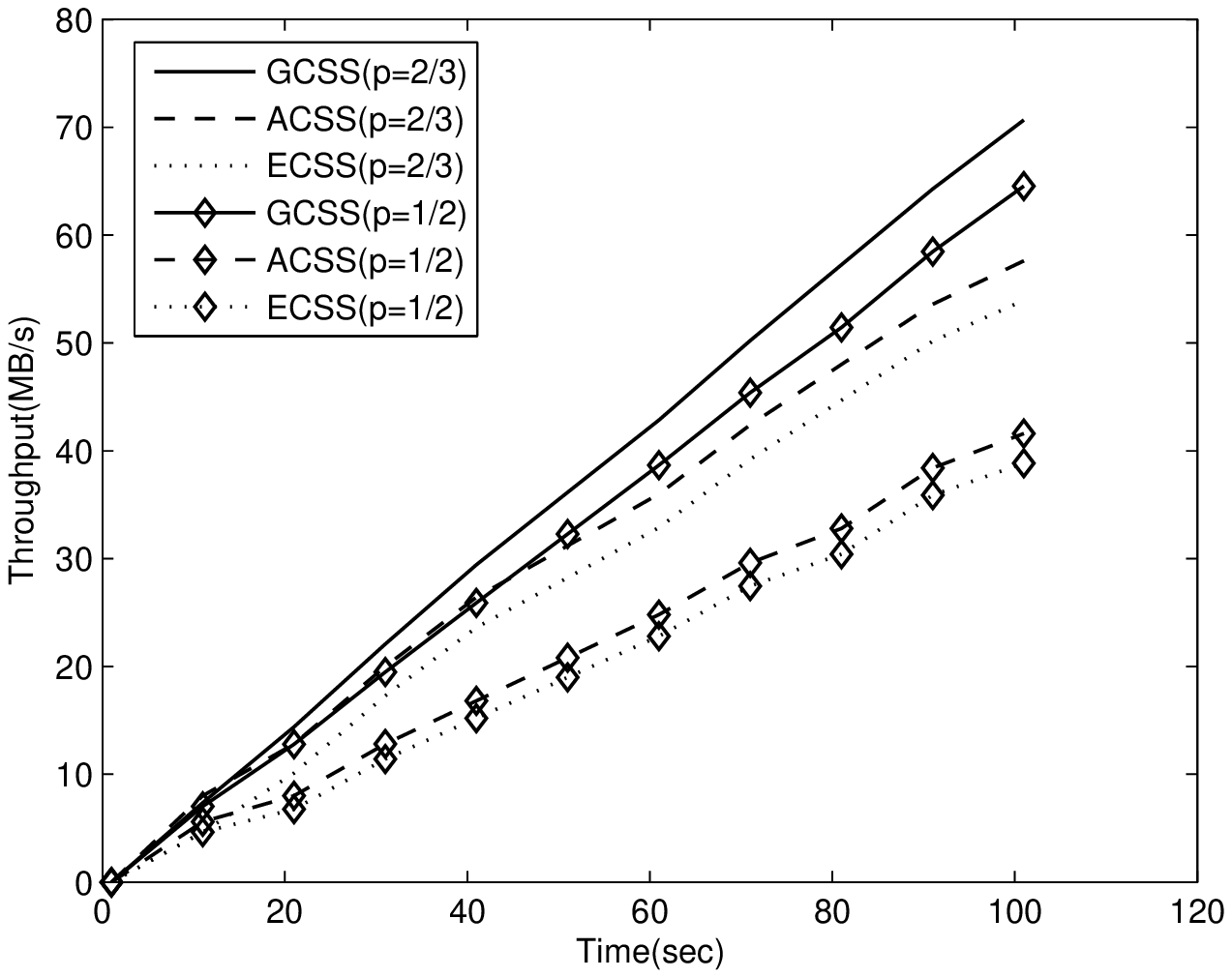}}
\caption{The non-saturation throughput in the time-invariant channel
case with different $p$.} \label{TH_INVUNSAT}
\end{figure}

Fig.\ref{TH_INVUNSAT} shows the non-saturation throughput comparison
among GCSS, ACSS and ECSS  in the time-invariant channel case when
$p=2/3$, $1/2$. Again, the $P_{f,th}=0.05$ is assumed as 0.05. It
can be observed that, GCSS substantially outperforms the other two
schemes. In addition, we notice that it will obtain higher
throughput if the channel availability $p$ becomes larger.

\begin{figure}[htb]
\centerline{\includegraphics[width=7.5cm]{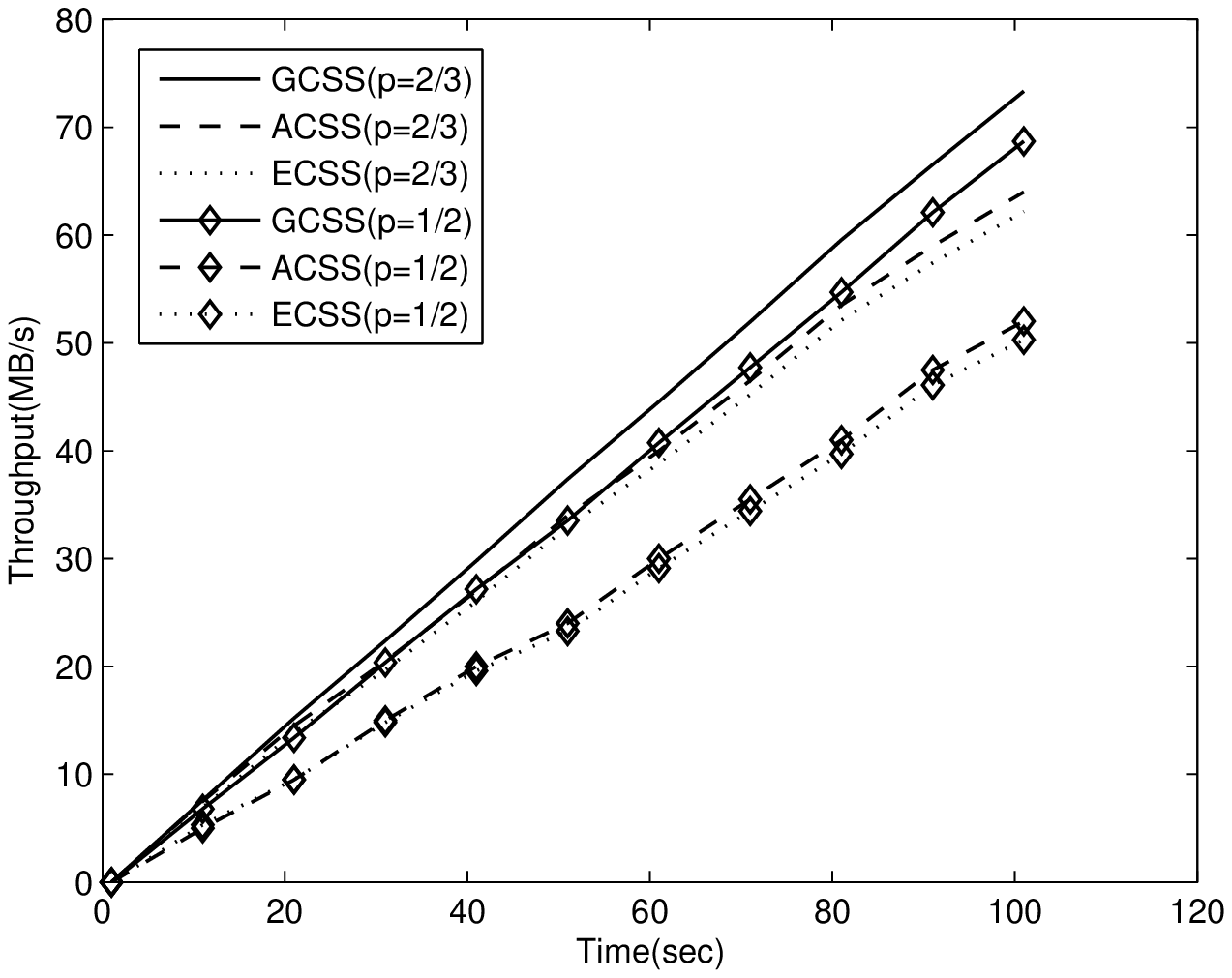}}
\caption{The saturation throughput in the time-varying channel case
with different $p$.} \label{TH_VARSAT}
\end{figure}

\begin{figure}[htb]
\centerline{\includegraphics[width=7.5cm]{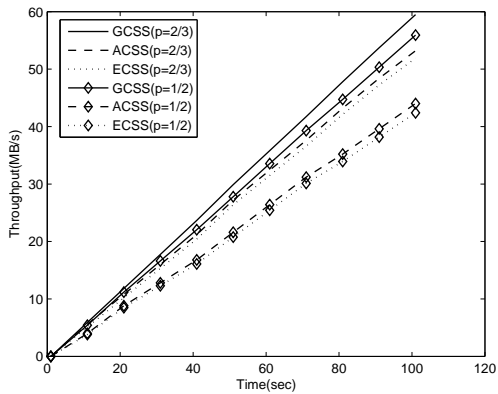}}
\caption{The non-saturation throughput in the time-varying channel
case with different $p$.} \label{TH_VARUNSAT}
\end{figure}

\subsubsection{Time-Varying Channel Case}
Fig.\ref{TH_VARSAT} and Fig.\ref{TH_VARUNSAT} show the saturation
and non-saturation throughput comparison among GCSS, ACSS and ECSS
in the time-varying channel case when $p=2/3$, $1/2$ and
$P_{f,th}=0.05$. The comparison indicates that GCSS is able to
achieve higher throughput than ACSS and ECSS. This is because GCSS
is able to detect and find more spectrum opportunities even when the
channel is dynamic. When the number of cooperative SUs becomes
larger, our scheme not only finds the available channel quicker but
also chooses the channel with maximal rate if more than one
available channels are found. Moreover, with the comparison to ECSS,
GCSS has the advantage of reducing sensing overhead. As a
consequence, the proposed GCSS achieves higher throughput in the
time-varying channel case.

\begin{figure}[htb]
\centerline{\includegraphics[width=7.5cm]{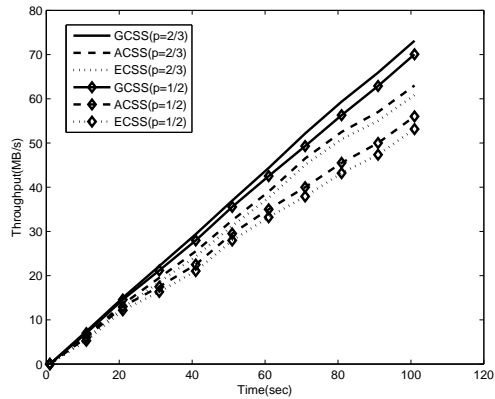}}
\caption{The saturation throughput comparison under complex-valued
signal system in the time-varying channel case with different $p$.}
\label{TH_VARSAT_Complex}
\end{figure}

\begin{figure}[htb]
\centerline{\includegraphics[width=7.5cm]{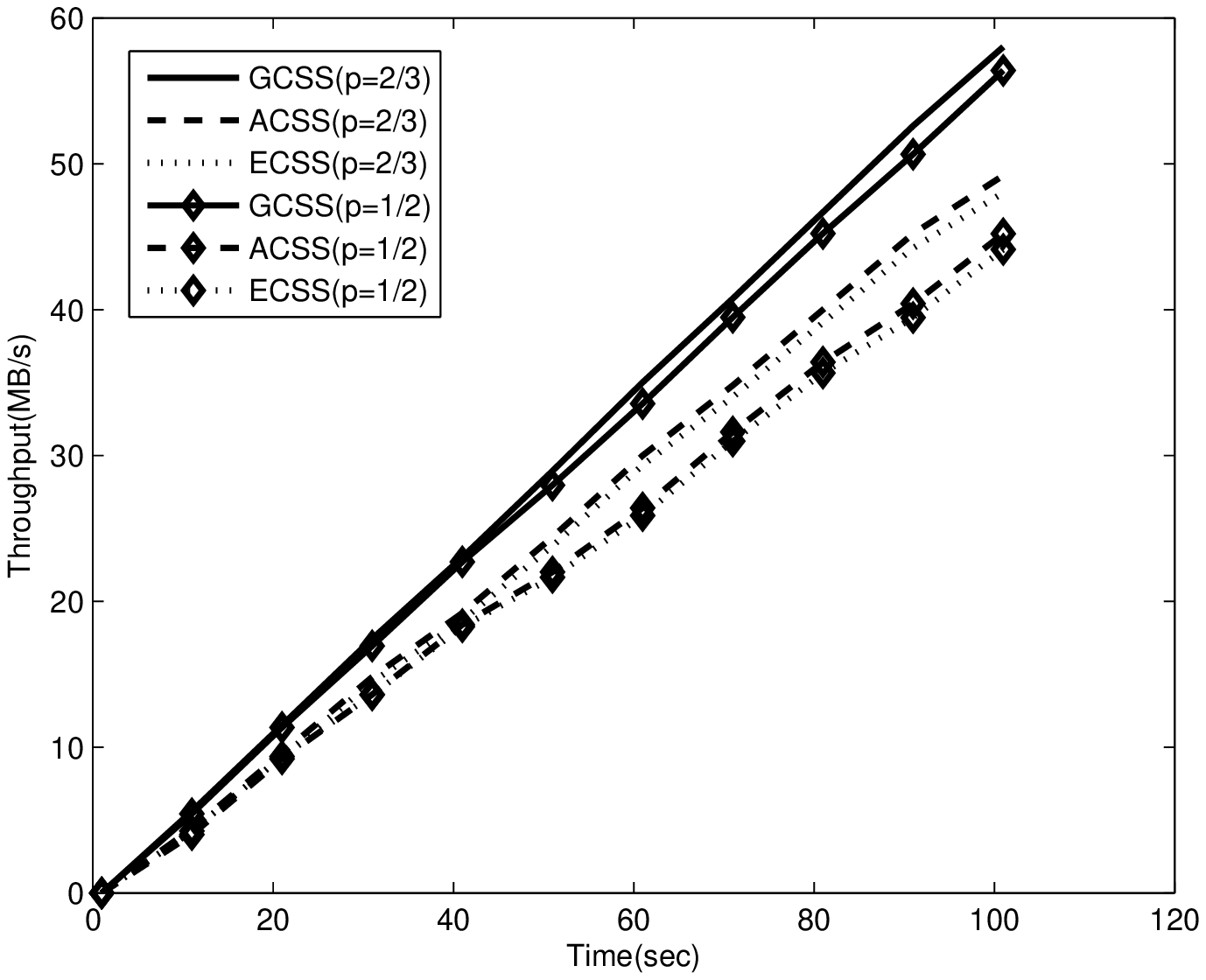}}
\caption{The  non-saturation throughput comparison under
complex-valued signal system in the time-varying channel case with
different $p$.} \label{TH_VARUNSAT_Complex}
\end{figure}

In addition, we illustrate the achievable throughput comparison
among GCSS, ACSS and ECSS under the complex-valued signal model.
Fig.\ref{TH_VARSAT_Complex} and Fig.\ref{TH_VARUNSAT_Complex} show
the saturation and non-saturation throughput comparison among GCSS,
ACSS and ECSS in the time-varying channel case, respectively. We
observe that GCSS also can obtain higher throughput than that in
ACSS and ECSS. This observation indicates the effectiveness of our
proposed MAC protocol in both of the real-valued and complex-valued
signal model.

\subsection{Sensing Overhead}

\begin{figure}[htb]
\centerline{\includegraphics[width=7.5cm]{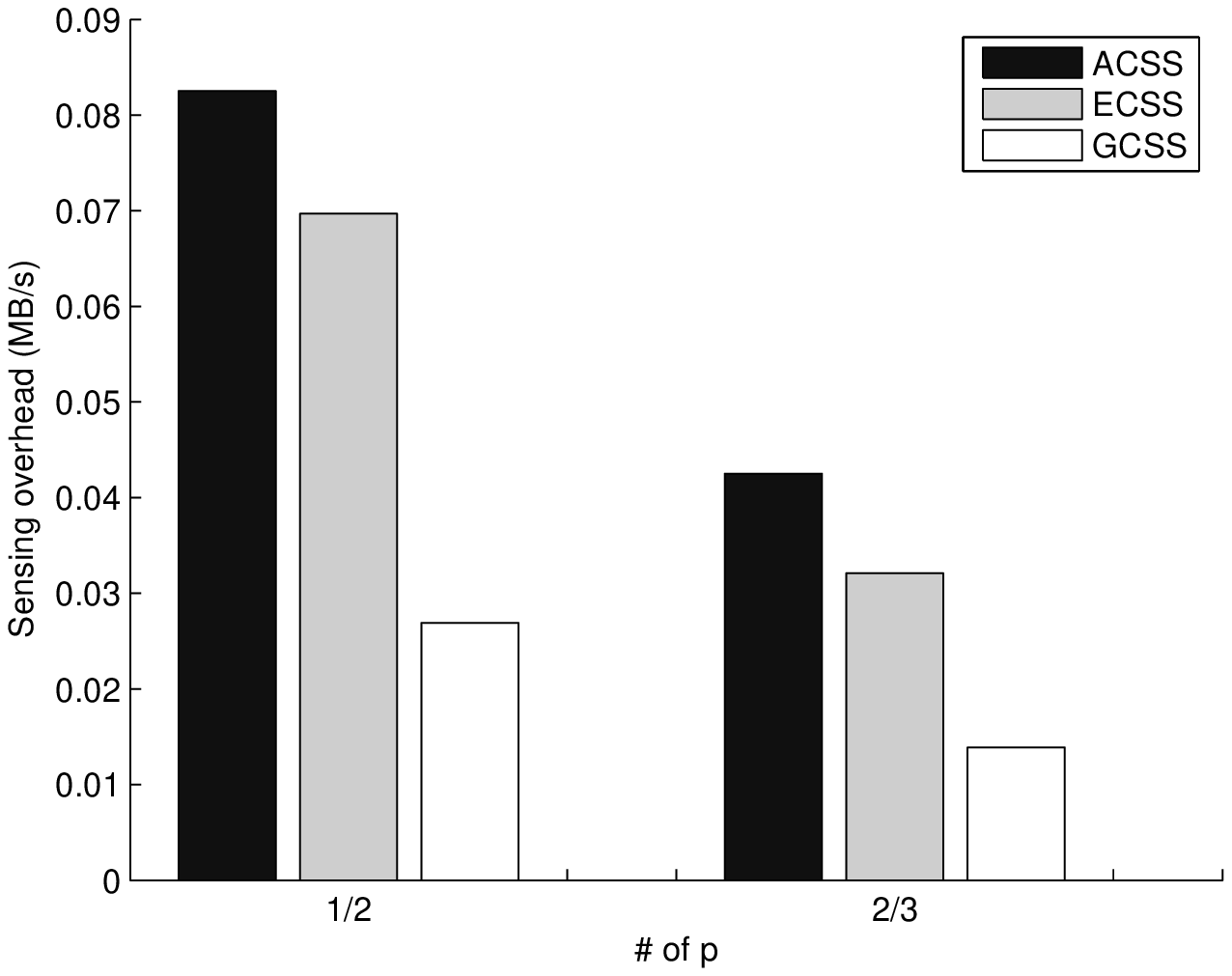}}
\caption{The sensing overhead in the time-invariant channel case
with different $p$ for saturation situation.} \label{SO_INVSAT}
\end{figure}

\begin{figure}[htb]
\centerline{\includegraphics[width=7.5cm]{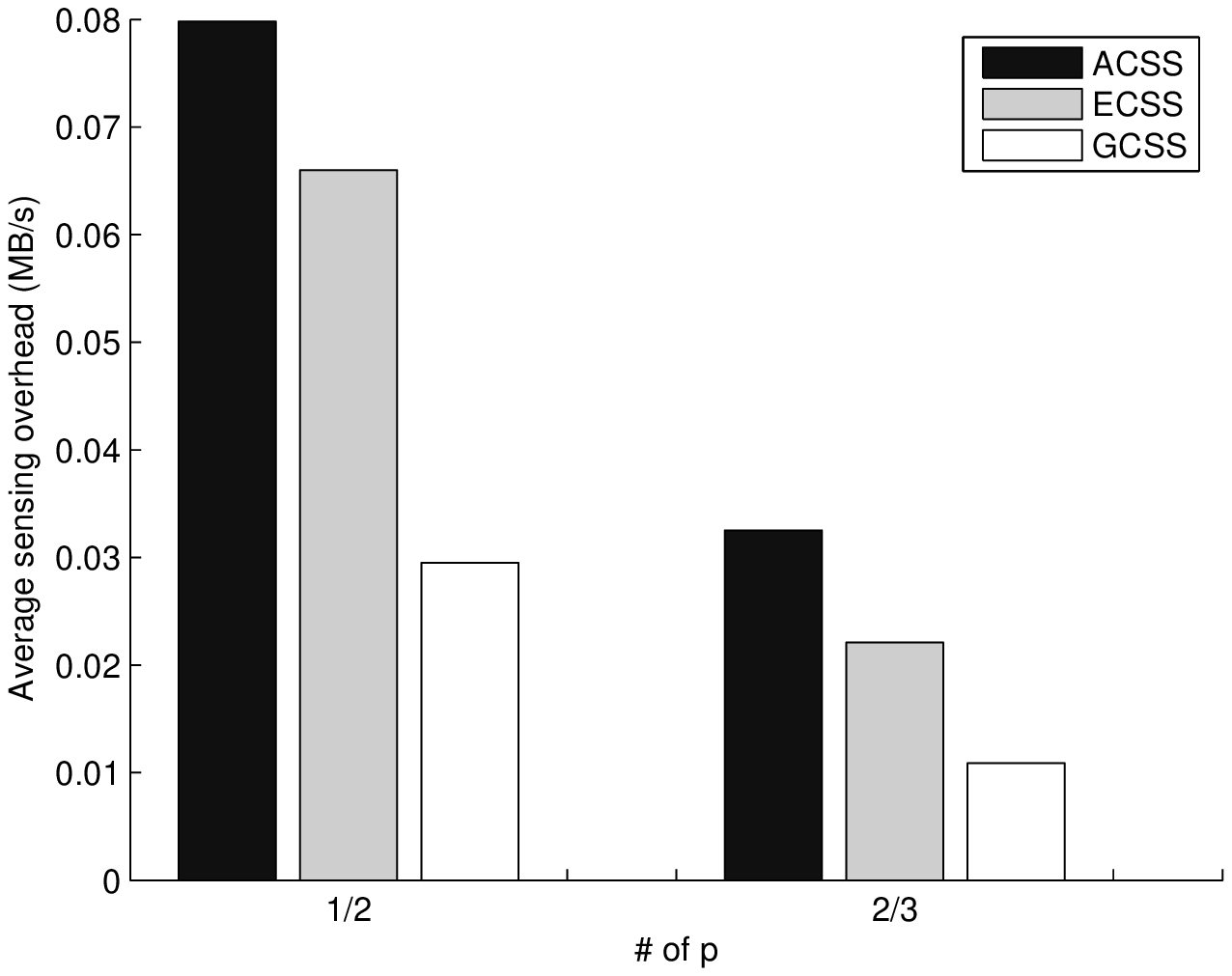}}
\caption{The sensing overhead in the time-invariant channel case
with different $p$ for non-saturation situation.}
\label{SO_INVUNSAT}
\end{figure}

\begin{figure}[htb]
\centerline{\includegraphics[width=7.5cm]{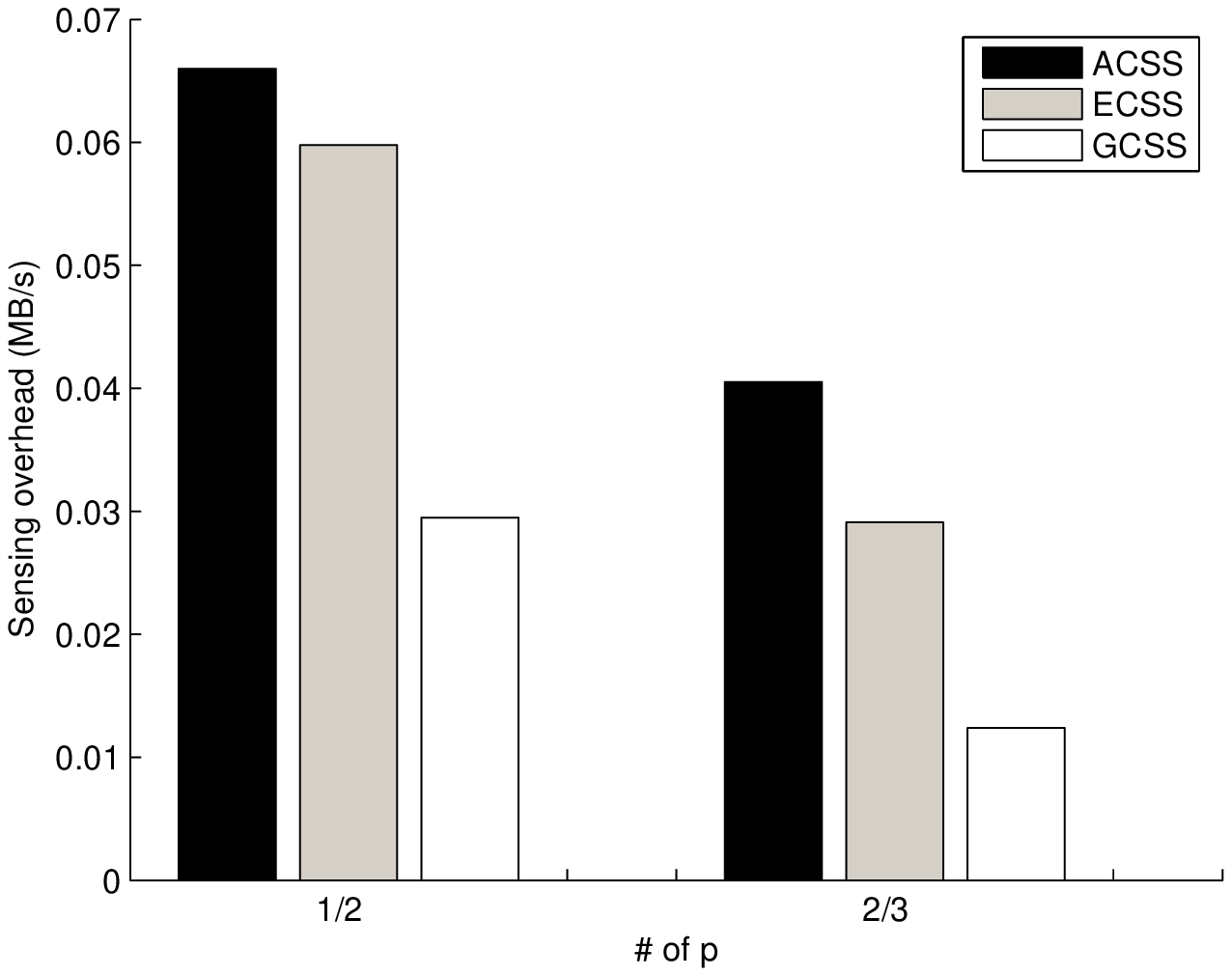}}
\caption{The sensing overhead in the time-varying channel case with
different $p$ for saturation situation.} \label{SO_VARSAT}
\end{figure}

\begin{figure}[htb]
\centerline{\includegraphics[width=7.5cm]{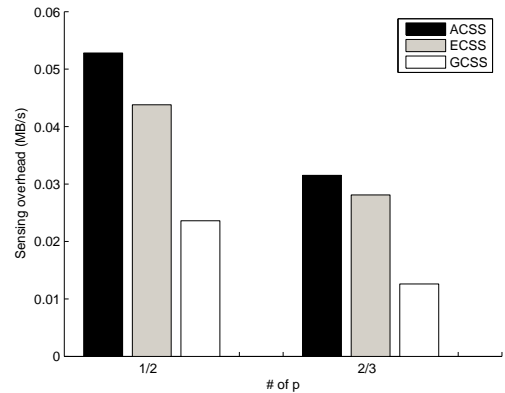}}
\caption{The sensing overhead in the time-varying channel case with
different $p$ for non-saturation situation.} \label{SO_VARUNSAT}
\end{figure}

\subsubsection{Time-Invariant Channel Case}
Fig.\ref{SO_INVSAT} shows sensing overhead among GCSS, ACSS and ECSS
in the time-invariant channel case for saturation situation. It is
observed that GCSS generates the lowest sensing overhead. This can
be explained as follows. GCSS selects the SUs to cooperate by using
the SU-selecting algorithm. The algorithm chooses the SUs with low
channel available probability ($P_{00}$) for the cooperative
sensing. This operation can substantially reduce sensing overhead by
avoiding the temporary stopping of the ongoing transmissions when
their channels are occupied by PUs. Comparatively, ACSS and ECSS
have no similar mechanisms and hence generate higher sensing
overhead. Fig.\ref{SO_INVUNSAT} shows the sensing overhead for
non-saturation situation. Similar observations and conclusions can
be made. In addition, we notice that sensing overhead decreases when
the channel availability $p$ becomes larger. With more channel
availability, there are more chances to find spectrum opportunities
in a fixed period; and hence less sensing overheads.

\subsubsection{Time-Varying Channel Case}

Considering the time-varying channel case, Fig.\ref{SO_VARSAT} and
Fig.\ref{SO_VARUNSAT} show the sensing overhead with different
channel availability $p$ under saturation and non-saturation
situation, respectively. It is clear that sensing overhead becomes
lower when the channel availability $p$ increases. Again, the
proposed GCSS incurs lower sensing overhead than ACSS and ECSS. With
the time-varying channel, we have considered the channel dynamics
and rate variation in selecting appropriate SUs to perform sensing.
Following this way, sensing overhead in traditional cooperative
sensing can be partially avoided.

\section{Conclusion}

We design an efficient MAC protocol with selective grouping and
cooperative sensing in cognitive radio networks. In our protocol,
the cooperative MAC can quickly discover the spectrum opportunities
without degrading sensing accuracy. An SU-selecting algorithm is
proposed for specifically choosing the cooperative SUs in order to
substantially reduce sensing overhead in both time-invariant and
time-varying channel cases. We formulate the throughput maximization
problems to determine the crucial design parameters and to
investigate the trade-off between sensing overhead and throughput.
Simulation results show that our proposed protocol can significantly
reduced sensing overhead without degrading sensing accuracy.

\begin{IEEEbiography}[{\includegraphics[width=1\textwidth]{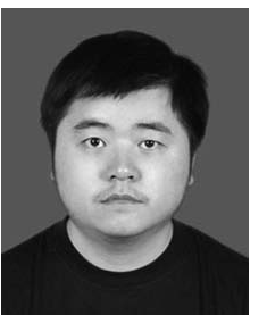}}]{Yi Liu}

 received his Ph.D. degree from South China University of Technology
(SCUT), Guangzhou, China, in 2011. After that, he worked in the
Institute of Intelligent Information Processing at Guangdong
University of Technology (GDUT). In 2011, he joined the Singapore
University of Technology and Design, where he is now a
post-doctoral. His research interests include cognitive radio
networks, cooperative communications, smart grid and intelligent
signal processing.
\end{IEEEbiography}

\begin{IEEEbiography}[{\includegraphics[width=1\textwidth]{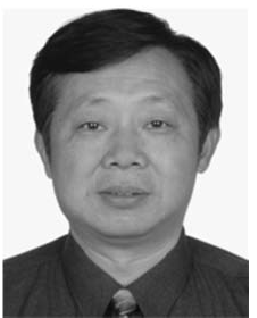}}]{Shengli Xie}

 [M'01-SM'02] received his M.S. degree in mathematics from Central China Normal University,
 Wuhan, China, in 1992 and his Ph.D. degree in control theory and applications from South
 China University of Technology, Guangzhou, China, in 1997.
He is presently a Full Professor with the Guangdong University of
Technology and a vice head of the Institute of Automation and Radio
Engineering. His research interests include automatic control and
blind signal processing. He is the author or coauthor of two books
and more than 70 scientific papers in journals and conference
proceedings. He received the second prize in China's State Natural
Science Award in 2009 for his work on blind source separation and
identification.

\end{IEEEbiography}

\begin{IEEEbiography}[{\includegraphics[width=1\textwidth]{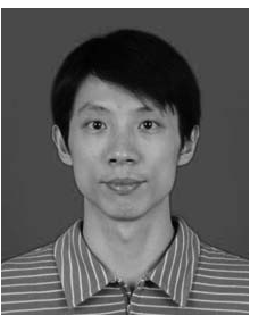}}]{Rong Yu}

 [S'05, M'08] (yurong@ieee.org) received his Ph.D. from Tsinghua University, Beijing, China, in 2007.
 After that, he worked in the School of Electronic and Information Engineering of
 South China University of Technology (SCUT). In 2010, he joined the Institute of Intelligent Information
 Processing at Guangdong University of Technology (GDUT), where he is now an associate professor.
 His research interest mainly focuses on wireless communications and networking, including cognitive radio,
 wireless sensor networks, and home networking. He is the co-inventor of over 10 patents and author or co-author
 of over 50 international journal and conference papers.

Dr. Yu is currently serving as the deputy secretary general of the
Internet of Things (IoT) Industry Alliance, Guangdong, China, and
the deputy head of the IoT Engineering Center, Guangdong, China. He
is the member of home networking standard committee in China, where
he leads the standardization work of three standards.

\end{IEEEbiography}

\begin{IEEEbiography}[{\includegraphics[width=1\textwidth]{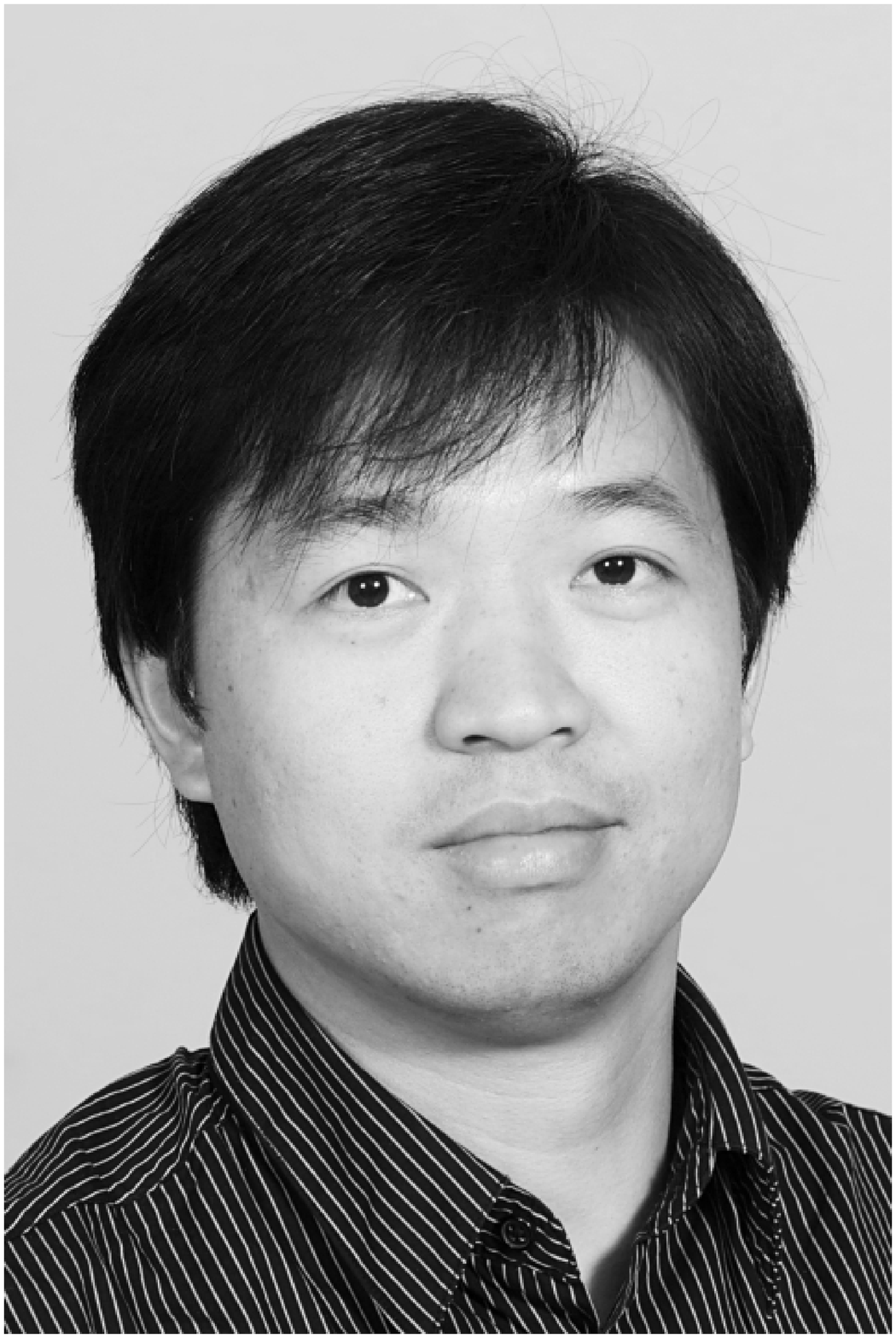}}]{Yan Zhang}

 (yanzhang@ieee.org) received a Ph.D. degree from Nanyang
Technological University, Singapore. He is working with Simula
Research Laboratory, Norway; and he is an adjunct Associate
Professor at the University of Oslo, Norway. He is an associate
editor or guest editor of a number of international journals. He
serves as organizing committee chairs for many international
conferences. His recent research interests include cognitive radio,
smart grid, and M2M communications.

\end{IEEEbiography}

\begin{IEEEbiography}[{\includegraphics[width=1\textwidth]{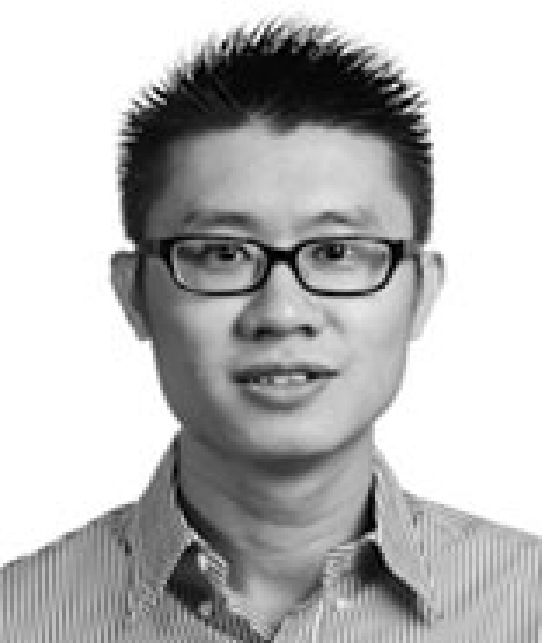}}]{Chau Yuen}

received the B. Eng and PhD degree from Nanyang Technological
University, Singapore in 2000 and 2004 respectively. Dr Yuen was a
Post Doc Fellow in Lucent Technologies Bell Labs, Murray Hill during
2005. He was a Visiting Assistant Professor of Hong Kong Polytechnic
University in 2008. During the period of 2006 - 2010, he worked at
the Institute for Infocomm Research (Singapore) as a Senior Research
Engineer. He joined Singapore University of Technology and Design as
an assistant professor from June 2010. He serves as an Associate
Editor for IEEE Transactions on Vehicular Technology. On 2012, he
received IEEE Asia-Pacific Outstanding Young Researcher Award.

\end{IEEEbiography}

%




\end{document}